\documentclass{article}
\usepackage{arxiv}
\pdfoutput=1

\usepackage[utf8]{inputenc} 
\usepackage[T1]{fontenc}    
\usepackage{hyperref}       
\usepackage{url}            
\usepackage{cite}
\usepackage{booktabs}       
\usepackage{amsfonts}       
\usepackage{nicefrac}       
\usepackage{microtype}      
\usepackage{lipsum}		
\usepackage{graphicx}
\usepackage{natbib}
\usepackage{doi}
\usepackage{threeparttable}
\usepackage{multirow} 
\usepackage{amsmath,amssymb,amsfonts}
\usepackage{algorithmic}
\usepackage{graphicx}
\usepackage{algorithm,algorithmic}
\usepackage{hyperref}
\usepackage{caption}
\usepackage{bm}
\usepackage{booktabs}
\usepackage{makecell}
\usepackage[inkscapeformat=png]{svg}
\usepackage[justification=centering]{caption}
\hypersetup{hidelinks=true}

\newcommand{\figref}[1]{Fig.~\ref{#1}}                      
\newcommand{\tabref}[1]{Table~\ref{#1}}                      

\newcommand{\tmdtp}{Densely-connected Temporal Pyramids}     
\newcommand{\dtpabb}{DTP-Net}

\title{DTP-Net: Learning to Reconstruct EEG signals in Time-Frequency Domain by Multi-scale Feature Reuse}

\author{ \href{https://orcid.org/0009-0001-8431-0422}{\includegraphics[scale=0.06]{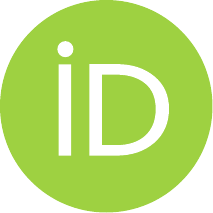}\hspace{1mm}Yan Pei} \\
	College of Biomedical Engineering \\
	Zhejiang University\\
	\texttt{summer\_bae@zju.edu.cn} \\
	\And
	{Jiahui Xu} \\
	Department of Neurology \\
        Sir Run Run Shaw Hospital \\
        School of Medicine \\
        Zhejiang University \\
	\texttt{xujiahui90@zju.edu.cn} \\
        \And
	{Qianhao Chen} \\
	College of Biomedical Engineering \\
	Zhejiang University\\
	\texttt{chen\_qh@zju.edu.cn} \\
        \And
	{Chenhao Wang} \\
	College of Biomedical Engineering \\
	Zhejiang University\\
	\texttt{12215047@zju.edu.cn} \\
        \And
	{Feng Yu*} \\
	College of Biomedical Engineering \\
	Zhejiang University\\
	\texttt{osfengyu@zju.edu.cn} \\
        \And
	{Lisan Zhang*} \\
	Department of Neurology \\
        Sir Run Run Shaw Hospital \\
        School of Medicine \\
        Zhejiang University \\
	\texttt{zls09@zju.edu.cn} \\
        \And
	{Wei Luo*} \\
	College of Biomedical Engineering \\
	Zhejiang University\\
	\texttt{luo.wei@zju.edu.cn} \\
}

\renewcommand{\shorttitle}{\textit{ }}

\hypersetup{
pdftitle={DTP-Net: Learning to Reconstruct EEG signals in Time-Frequency Domain by Multi-scale Feature Reuse},
pdfsubject={q-bio.NC, q-bio.QM},
pdfauthor={Yan Pei, Jiahui Xu, Qianhao Chen, Chenhao Wang, Feng Yu*, Lisan Zhang* and Wei Luo*},
pdfkeywords={Electroencephalogram (EEG), deep learning, fully convolutional neural network, artifact removal, signal reconstruction, Time-Frequency domain},
}

\begin{document}
\maketitle

\begin{abstract}
Electroencephalography (EEG) signals are easily corrupted by various artifacts, making artifact removal crucial for improving signal quality in scenarios such as disease diagnosis and brain-computer interface (BCI). In this paper, we present a fully convolutional neural architecture, called DTP-Net, which consists of a Densely Connected Temporal Pyramid (DTP) sandwiched between a pair of learnable time-frequency transformations for end-to-end electroencephalogram (EEG) denoising. 
The proposed method first transforms a single-channel EEG signal of arbitrary length into the time-frequency domain via an Encoder layer. 
Then, noises, such as ocular and muscle artifacts, are extracted by DTP in a multi-scale fashion and reduced. 
Finally, a Decoder layer is employed to reconstruct the artifact-reduced EEG signal. 
Additionally, we conduct an in-depth analysis of the representation learning behavior of each module in DTP-Net to substantiate its robustness and reliability.
Extensive experiments conducted on two public semi-simulated datasets demonstrate the effective artifact removal performance of DTP-Net, which outperforms state-of-art approaches. Experimental results demonstrate cleaner waveforms and significant improvement in Signal-to-Noise Ratio (SNR) and Relative Root Mean Square Error (RRMSE) after denoised by the proposed model.
Moreover, the proposed DTP-Net is applied in a specific BCI downstream task, improving the classification accuracy by up to 5.55\% compared to that of the raw signals, validating its potential applications in the fields of EEG-based neuroscience and neuro-engineering.
\end{abstract}


\section{Introduction}
\label{sec:introduction}
Electroencephalography(EEG) provides a non-invasive way to measure brain activity with high temporal resolution, but it can be easily contaminated by various artifacts, including those generated by muscle and eye movement. Removing such interferences with minimum distortion would improve the signal-to-inference/artifacts ratio (SIR/SAR) of the recordings and enhance the performance of downstream EEG interpretation tasks, such as sleep data analysis and brain-computer interface (BCI) applications \citep{chen:2022:eegdl}.

Blind source separation (BSS) approaches provide a straightforward way to denoise EEG by factorizing contaminated recordings into signal and artifact components. Among these methods, independent component analysis (ICA) assumes that the sources in an EEG recording are statistically independent. This can make it difficult to extract artifacts precisely without distorting the actual neuronal signal if the assumption does not hold \citep{jung:1998:ica_eeg}. Canonical correlation analysis (CCA) is another BSS method, which separates EEG artifacts based on a weaker assumption that sources are uncorrelated but autocorrelated \citep{wim:2006:cca_eeg}. However, these methods are unable to remove artifacts when recordings from only one EEG channel are available \citep{mumtaz:2021:eeg_denoise_review}. To address this, single-channel BSS methods, such as  wavelet-ICA \citep{Inuso2007}, EMD-ICA \citep{mijovic:2010:emd-ica}, and EEMD-CCA \citep{sweeney:2013:emd-cca}, were proposed. The common idea behind these hybrid methods is to decompose an EEG signal into components with disjoint spectra using wavelet transformation or empirical mode decomposition (EMD). Note that in wavelet-based BSS methods, it can be difficult to choose the mother wavelet without prior knowledge. In reality, BSS techniques often rely on the assumption of linear combinations of components, which may be invalid when nonlinearities exist within the recording system \citep{mannan:2018:denoise_review}. 
\setlength{\arrayrulewidth}{0.3mm}
\renewcommand{\arraystretch}{1.5}
\begin{table*}
\caption{Comparison of Deep Learning-based EEG Artifact Removal Algorithms in Literature}
\label{tab:method_review}
    \begin{center}
    \resizebox{\linewidth}{!}{
        \begin{tabular}{c c c c c c c}
        \hline
        Study & Dataset Used & Evaluation Metrics& Techniques Used & \makecell[c]{Applicable to \\ Single EEG Channel} & Artifact Type & \makecell[c]{Applicable to EEG Signals \\ of Arbitrary Length}\\
        \specialrule{0.05em}{3pt}{3pt}
         \citep{yang:2018:dleegdenoise} & 1. BCI Competition Dataset  \citep{blankertz:2007:bci_dataset} &\makecell[c]{ PSD, RRMSE,\\  EEG Classification Accuracy} & Stacked Sparse Autoencoder (SSAE) & Yes & EOG & No \\
        \specialrule{0em}{3pt}{3pt}
          \citep{sun:2020:dleegdenoise} &\makecell[c]{ 1. CHB-MIT Dataset (\href{http://www.physionet.org}{http://www.physionet.org}) ; \\  2. Noise Dataset (\href{http://www.physionet.org}{http://www.physionet.org})}  & \makecell[c]{SNR, RRMSE, ApEn, \\ Correlation Coeeficient (CC)} & Inception-ResNet & Yes & EMG, ECG, EOG & No \\
         \specialrule{0em}{3pt}{3pt}
          \citep{zhang:2021:cnndenoise} & 1. EEGDenoiseNet  \citep{zhang:2021:eegdenoisenet} & RRMSE, CC & \makecell[c]{Feedforward Convolutional \\ Neural Networks (CNNs)} & Yes & EMG & No\\
         \specialrule{0em}{3pt}{3pt}
          \citep{swaangjai:2022:eeganet} & \makecell[c]{1. EEG Eye Artifact Dataset;\\ 2. BCI Competition IV 2b  \citep{tangermann:2012:bci_competition};\\ 3. Multimodal Signal Dataset  \citep{jeong:2020:multimodal_dataset}} & \makecell[c]{RRMSE, CC, SNR, \\ EEG Classification Accuracy} & Generative Adversarial Network (GAN) & Yes & EOG & No\\
         \specialrule{0em}{3pt}{3pt}
          \citep{zhang:2022:multimodule_denoise} & EEGDenoiseNet  \citep{zhang:2021:eegdenoisenet} & RRMSE, CC & \makecell[c]{Multi-Module Neural \\ Network (MMNN)} & Yes & EMG, EOG & No\\
         \specialrule{0em}{3pt}{3pt}
          \citep{Gao2022} & \makecell[c]{1. Synthesised Contaminated EEG Dataset; \\ 2. KLADOS Dataset  \citep{klados:2016:eeg_eog_dataset}} & SNR, RRMSE, CC & \makecell[c]{CNN,  \\ Long Short-Term Memory (LSTM)} & Yes & EMG, EOG & No \\
         \specialrule{0em}{3pt}{3pt}
          \citep{chuang:2022:icunet} & \makecell[c]{1. Resting-state EEG Dataset; \\ 2. Simulated-Driving Dataset  \citep{chuang:2014:driving_task_dataset} \\ 3. BCI Challenge Dataset  \citep{margaux:2012:bci_ner}} & \makecell[c]{MSE, SNR, \\ EEG Classification Accuracy} & ICA, U-Net & No & \makecell[c]{EMG, EOG \\ Channel Artifacts, etc.} & Yes\\
         \hline
        \end{tabular}
    }
    \end{center}
\end{table*}

On the other hand, deep neural networks can learn to denoise EEG signals without relying on restrictive assumptions.  Autoencoders \citep{yang:2018:dleegdenoise}, convolutional neural networks (CNNs) \citep{sun:2020:dleegdenoise} \citep{zhang:2021:cnndenoise} \citep{zhang:2022:multimodule_denoise} \citep{zhang:2021:eegdenoisenet}, generative Adversarial Networks (GANs) \citep{swaangjai:2022:eeganet}, recurrent neural networks (RNNs) \citep{chuang:2022:icunet} and combination of them \citep{Gao2022} have been recently proposed. For example, Yang et al. \citep{yang:2018:dleegdenoise} applied the stacked sparse autoencoder \citep{vincent:2010:ssae} (SSAE) consisting of multiple layers of sparse autoencoders and combined with greedy layer-wise training to skillfully remove ocular artifacts. Sun et al.  \citep{sun:2020:dleegdenoise} proposed a new one-dimensional residual convolutional neural network (1D-ResCNN) which utilized the structural features of CNN and Inception-ResNet to extract complex nonlinear features of EEG signals for artifact removal, making it the first CNN-based approach for EEG denoising. Zhang et al. \citep{zhang:2021:cnndenoise} introduced a novel CNN with gradually ascending feature dimensions and downsampling in time series to remove muscle artifacts in EEG data, and the proposed model achieves performance improvement in SNR and RRMSE. Sawangjai et al.  \citep{swaangjai:2022:eeganet} proposed a framework, EEGANet, to eliminate ocular artifacts from contaminated EEG, adapting from SRGAN \citep{ledig:2017:srgan} which was optimized to generate super-resolution images. This is the first application of GAN in EEG artifact reduction. Zhang et al. \citep{zhang:2022:multimodule_denoise} built a multi-module neural network (MMNN) to remove multiple types of artifacts from noisy EEG signals, consisting of multiple denoising modules connected in parallel. Each module is built upon one-dimensional convolutions and fully connected layers which act as an end-to-end trainable filter. However, these approaches utilize off-the-shelf deep learning models without considering the non-stationarity and complex characteristics of EEG signals.

Brain characteristics are dynamic and can exhibit variations in time. We argue that multi-resolution analysis which extract the representation at different scales to adapt to the varying dynamics of EEG signals facilitates artifacts removal task \citep{clark:1995:eeg_multiresolution}. To this end, multi-scale context information should be propagated across neural layers in an appropriate domain for this signal-to-signal translation task. Gao et al. proposed dual-scale CNN-LSTM (Long Short-Term Memory), in which two CNN branches with different kernel sizes are utilized to construct morphological features of different scales and the temporal dependencies are autonomously learned by LSTM \citep{Gao2022}. Nonetheless, dual-scale deep features extracted by the dual-branch model is inadequate to deal with intricate and dynamic EEG signals. Chuang et al. developed a new model, IC-U-Net, by combining the strengths of ICA and U-Net \citep{ronneberger:2015:unet} in EEG artifact removal \citep{chuang:2022:icunet}. This model was trained with mixtures of independent EEG independent components (ICs), based on the U-Net model with a loss function ensemble. The proposed IC-U-Net can learn multi-scale deep features from input signals through its downsample encoder and upsample decoder architectures and skip connections, respectively. Whereas, stacking non-local operators in the encoder and decoder may impair network efficiency and cause the loss of the spatial information  \citep{wang:2020:non-local-unet}. The comparison of deep learning-based EEG artifacts removal algorithms is listed in \tabref{tab:method_review}

\begin{figure*}
    \centering
    \includegraphics[width=1.0\textwidth]{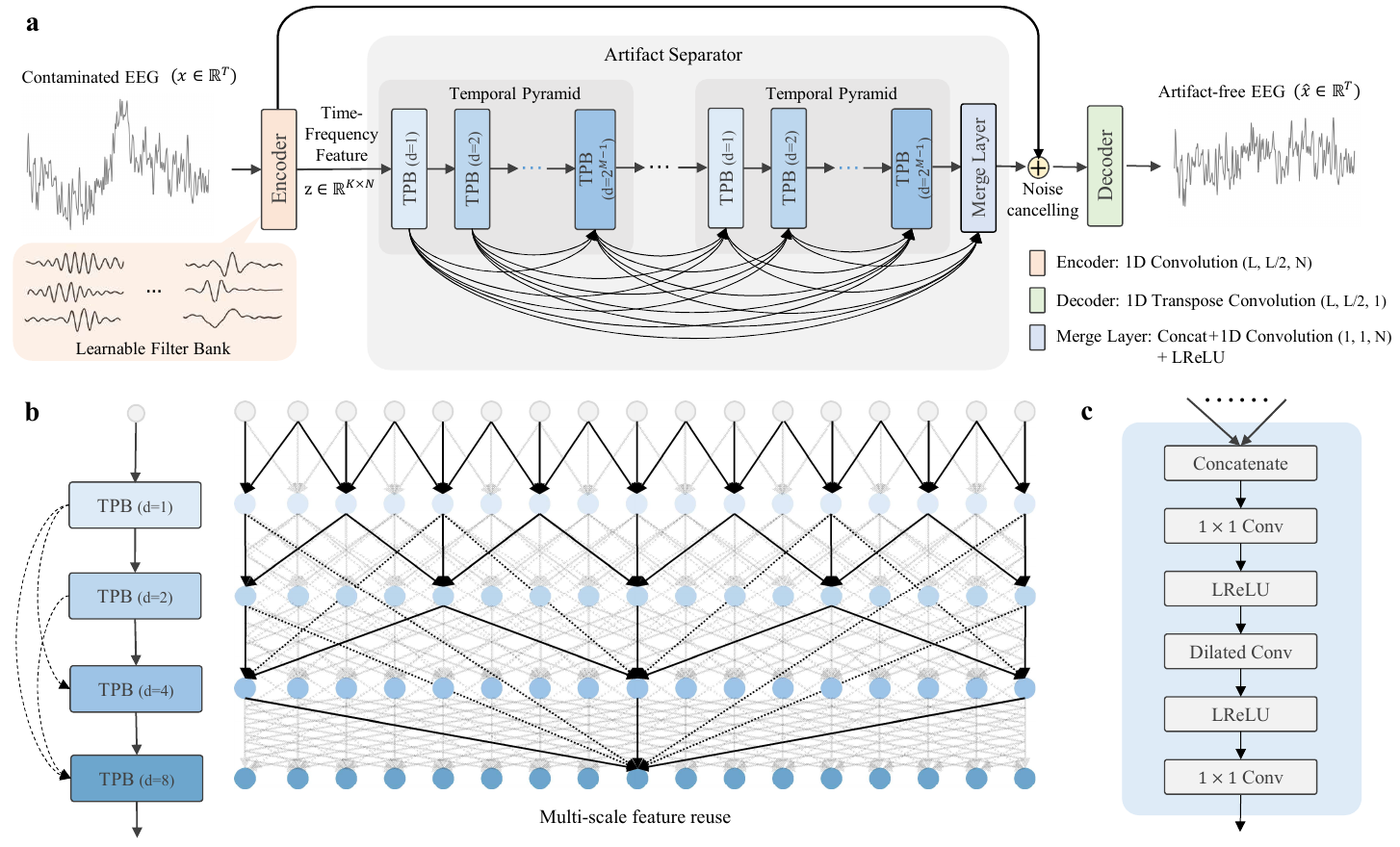}
    \caption{The macro-architecture and micro-structures of the proposed \dtpabb. 
    (a) The overall architecture. 
    (b) The structure of an example Temporal Pyramid with $M=4$ and the illustration of multi-scale feature reuse. Note that shortcut links from the preceding Temporal Pyramid are not depicted for simplicity. 
    (c) The structure of each building block in Temporal Pyramids.}
    \label{fig:overall}
\end{figure*}

We argue that successful artifact removal requires the network to gather global information by capturing long-term dependencies within the recordings, while maintaining the local details of signals. To this end, multi-scale context information should be propagated across neural layers in an appropriate domain for this signal-to-signal translation task. Gao et al. \citep{Gao2022} proposed dual-scale CNN-LSTM (Long Short-Term Memory), in which two CNN branches with different kernel sizes are constructed to extract dual-scale morphological features of contaminated EEG signals. Nonetheless, features of different scales are learned separately. Hence, the contextual dependencies between global and local information are ignored. Chuang et al. combined ICA and U-Net  \citep{ronneberger:2015:unet} in EEG artifact removal  \citep{chuang:2022:icunet}. The proposed IC-U-Net can learn both global and local contextual features from input signals through its Encoder-Decoder architecture and skip connections, respectively. Whereas stacking non-local operators in the Encoder and Decoder may impair network efficiency and cause the loss of the spatial information  \citep{wang:2020:non-local-unet}. Using dilated convolutions  \citep{yu:2015:dilate} is another way to achieve multi-scale context aggregation. Based on this, Luo et al. proposed mixed-context networks (MCN) in which each layer can simultaneously process local and global contextual information for image-to-image translation tasks  \citep{luo:2022:dlpr}. Yet, the back-propagation of gradients during training is impeded due to its simple cascaded architecture.

In this paper, we propose a novel fully-convolutional neural network model, namely, \tmdtp \ network (\dtpabb), for artifacts removal of single-channel EEG. The macro-architecture of \dtpabb \ is shown in \figref{fig:overall}(a), in which a pair of Encoder and Decoder transitions the input signal to and back from the temporal-frequency domain where artifacts are separated, respectively. The micro-structures inside the artifact separator mainly include convolutional modules with different dilation rates, as shown in \figref{fig:overall}(b). These modules are densely connected to facilitate multi-scale feature reuse and easier gradient back-propagation. Note that the proposed \dtpabb \ can denoise EEG recordings of any length thanks to its fully-convolutional structure, while in networks with fully connected layers, the length of the network input is pre-fixed. The main contributions of this work are as follows:
\begin{itemize}
    \item[1)] 
     We propose \dtpabb \ architecture, which can be used to separate out artifacts from recordings of any length in the time-frequency domain. The effectiveness of \dtpabb \ model in denoising EEG signals is validated on two semi-synthetic datasets. Additionally, we demonstrate the feasibility of \dtpabb \ model in downstream applications by conducting additional experiments on the BCI dataset.
     \item[2)] 
     We investigate the underlying mechanisms and dynamics of the \dtpabb \ model by analyzing the representation learning behavior of Encoder layer and artifact separator. We empirically observe the mechanism for the Encoder layer to learn distinguishable time-frequency features from input signals and the artifact separator extracting multi-scale artifact features through periodically hierarchical representation learning. 
     \item[3)] 
     We provide a ready-to-use EEG-denoise tool with a pre-trained \dtpabb \ model, which can be found at https://github.com/WilliamRo/EEG-Denoise.
\end{itemize}


\section{METHOD}

\label{sec:method}
In this section, we detail the proposed EEG signal denoising model, \dtpabb \, which mainly consists of an Encoder, an artifact separator, and a Decoder, as shown in \figref{fig:overall}. The Encoder layer is first used to transform the input signal $x$ into its time-frequency representation $z$. In the time-frequency domain, artifact components are extracted by an artifact separator comprised of a series of densely connected Temporal Pyramid Blocks (TPBs), which can aggregate both local and global context information. The noise-reduced EEG signal $\hat{x}$ is then reconstructed by the Decoder from its time-frequency representation. The whole denoising process can be formulated as follows:
\begin{equation}
    z = \text{Encoder}(x),
\end{equation}
\begin{equation}
    \hat{x} = \text{Decoder}(z + \text{Separator}(z)).
\end{equation}
The macro-architecture and micro-structures of the proposed \dtpabb \ model are detailed in the following subsections.

\subsection{Encoder and Decoder}
The Encoder in \dtpabb \ extracts features that represent how the frequency content of the non-stationary EEG signal changes over time. Different from the short-time Fourier transform (STFT), which uses a set of sinusoidal curves with linearly increasing frequencies as bases, we allow the filter bank to be learned adaptively by observing pairs of contaminated and artifact-free EEG signals. This is achieved by using an $N$-channel 1D convolutional layer with a kernel size of $L$ and a stride size of $\frac{L}{2}$. In this way, an input signal $x\in \mathbb{R}^T $ is windowed into $K$ half-overlapped frames for each channel. We denote the output of the Encoder as $z\in \mathbb{R}^{K\times N}$. Here $K = \frac{2T}{L}-1$, and $T$ denotes the signal length.

The trainable parameters in the Encoder $W \in \mathbb{R}^{N \times L}$ can be thought of as a finite impulse-response filter bank. In Section IV, we introduce a different perspective on the Frequency Principle for neural network models used in signal transition tasks by analyzing $W$ during training.

Corresponding to the Encoder, we use a single-channel transposed convolutional layer as the Decoder, which is also trainable. The Decoder transforms an artifact-reduced signal from a time-frequency representation $\hat{z}\in \mathbb{R}^{K\times N}$ back to a time-domain representation $\hat{x} \in \mathbb{R}^{T}$. Note that we do not use any nonlinear activation following the Encoder/Decoder operation. Otherwise, the original signal is unlikely to be recoverable by the Decoder using its time-frequency representation calculated by the Encoder, i.e., the following property does not hold:

\begin{equation}
    x=\text{Decoder}(\text{Encoder}(x)).
\end{equation}
We argue that this is necessary, since a denoiser should not distort a noisy-free signal, i.e., $\hat{x}= \text{Denoiser}(\hat{x})$.

\subsection{Artifact seperator}
The artifact separator in \figref{fig:overall}(a) extracts the time-frequency components of artifacts. It incorporates a cascade of $R$ Temporal Pyramids followed by a feature merge layer. As shown in \figref{fig:overall}(b), each Temporal Pyramid contains $M$ Temporal Pyramid Blocks (TPBs) with exponentially increased dilation rates $1, 2, 4, \cdots, 2^{M-1}$, respectively. Besides, each block receives the output features from all preceding blocks. In other words, those TPBs are densely connected to ensure maximum information flow  \citep{huang2017densely}, enabling multi-scale feature reuse and strengthening feature propagation.  At the end of the separator, artifact features from different scales are merged in order to mitigate them from the time-frequency representation of the contaminated signal.

The advantages of such a multi-scale densely connected architecture have two folds. First, residual links allow multi-scale contextual information extracted from different receptive fields to be aggregated in each TPB. In other words, each TPB can process local and global features simultaneously, which is crucial for successful dense prediction tasks  \citep{luo:2022:dlpr}. Second, shortcut connections can create a strong dependency between network input and output, allowing for the faithful recovery of detailed signal structures. Meanwhile, for each TPB in the network, the direct access to all the preceding features and the gradients from the final loss function allows implicit deep supervision  \citep{lee:2015:deeply_supervised_nets}, alleviating the vanishing and exploding gradient problem.

\begin{figure}
    \centering
    \includegraphics[width=0.5\textwidth]{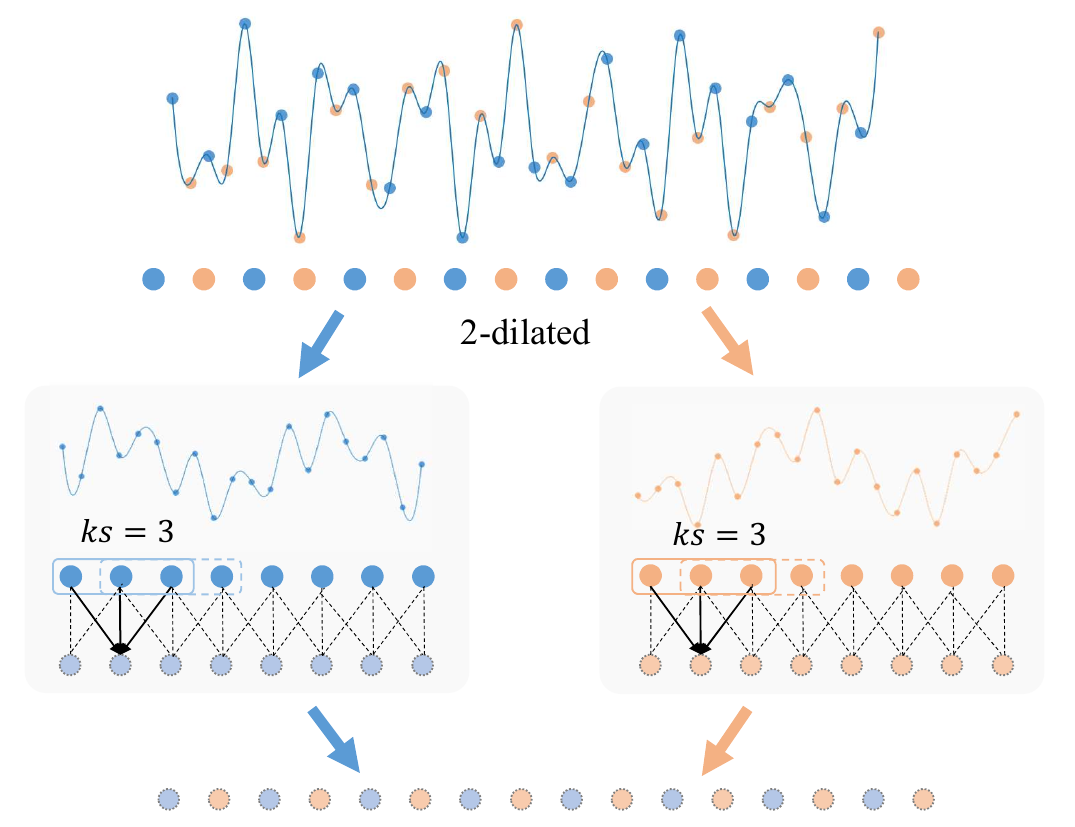}
    \caption{Illustration of 1D 2-dilated convolution in the context of signal processing. In general, applying 1D $d$-dilated convolution to a signal with a sampling frequency of $f$ can be seen as convolving $d$ signals with a sampling frequency of $\frac{f}{d}$ and then assembling the results in a time-interleaved fashion.}
    \label{fig:dilation}
\end{figure}

\subsection{Temporal Pyramid Block (TPB)}
The receptive field of each TPB is determined by the dilated convolution layer  \citep{yu:2015:dilate} in its micro structure to ensure a suitable large context window that can capitalize on the long-range dependency of time series signal. As shown in \figref{fig:overall}(c), multi-scale features from all preceding blocks are first concatenated channel-wisely and then fed into a bottleneck layer, which has the same structure as the last layer in the TPB structure. These bottleneck layers are 1D convolutional layers with a kernel size of 1 and a stride of 1. They have relatively fewer channels compared to dilated convolutional layers in order to reduce the number of trainable parameters. Note that we do not use element-wise parametric activation functions, such as PReLU  \citep{He2015}, to avoid a diversified process across different temporal locations. Furthermore, this type of activation function requires fixed-length inputs for neural networks.

 Moreover from the perspective of digital filtering, a $d$-dilated convolution can be viewed as convolving $d$ signals segments of signals that have been downsampled by a factor of $d$. In this way, high-frequency features can be assembled by interleaving low-frequency features that are easier for a neural network to extract  \citep{xu:2019:train_behaviour}, as illustrated in \figref{fig:dilation}.



\section{EXPERIMENTS AND RESULTS}

\subsection{Datasets}
To validate the efficiency and feasibility of the proposed method, we evaluate its performance on three public datasets, including one single-channel dataset EEGDenoiseNet  \citep{zhang:2021:eegdenoisenet}, one multi-channel dataset Semi-Simulated EEG/EOG  \citep{klados:2016:eeg_eog_dataset}, and one downstream BCI dataset MNE M/EEG. The detailed information is described as follows:

\setlength{\arrayrulewidth}{0.4mm}
\renewcommand{\arraystretch}{1.5}
\begin{table*}
\caption{Architecture Details of \dtpabb s in Different Tasks}
\label{tab:hp}
    \begin{center}
        \begin{tabular}{c c c c c c c c c }
        \hline
        Dataset & Artifact & Sample Rate (Hz) & N & L & H & P & M & R\\
        \hline
        \multirow{3}*{EEG DenoiseNet} & EMG & 512 & 454 & 23 & 440 & 5 & 5 & 7 \\
        & EOG & 256 & 248 & 32 & 394 & 4 & 6 & 6 \\
        & EMG+EOG & 256 & 305 & 32 & 243 & 3 & 5 & 5\\
        \hline
        Semi-Simulated & EOG & 200 & 512 & 8 & 64 & 3 & 6 & 4\\
        \hline
        \end{tabular}
    \end{center}
\end{table*}

\setlength{\arrayrulewidth}{0.4mm}
\renewcommand{\arraystretch}{1.7}
\begin{table*}
\caption{Comparison of Different Model Complexities}
\label{tab:tp}
    \begin{center}
    \resizebox{\linewidth}{!}{
        \begin{tabular}{c c c c c c c | c c c c c}
        \toprule[2pt]
        \multirow{2}*{\textbf{Dataset}} & \multirow{2}*{\textbf{Artifact}} & \multicolumn{5}{c}{\textbf{The Number of Trainable Parameters}} & \multicolumn{5}{c}{\textbf{Floating Point Operations (FLOPs)}} \\ \cline{3-12}
         {} & {} &\textbf{1D-ResCNN} & \textbf{SimpleCNN} & \textbf{NovelCNN} & \textbf{WQN} & \textbf{\dtpabb \ (Ours)} & \textbf{1D-ResCNN} & \textbf{SimpleCNN} & \textbf{NovelCNN} & \textbf{WQN} & \textbf{\dtpabb \ (Ours)} \\
        \midrule[2pt]
        \multirow{3}*{EEG DenoiseNet} & EMG & 33.6M & 67.1M & 58.7M & - & 45.6M & 67.2M & 134.3M & 117.4M & - & 91.2M\\
        {}& EOG & 8.4M & 16.8M & 33.5M & - & 39.8M & 16.9M & 33.6M & 67.1M & - & 79.6M\\
        {}& EMG+EOG & 8.4M & 16.8M & 33.5M & - & 10.0M & 16.9M & 33.6M & 67.1M & - & 20.1M\\
        \hline
        Semi-Simulated & EOG & 9.4M & 18.7M & 34.0M & - & 2.4M & 18.8M & 37.4M & 70.2M & - & 4.7M\\
        \hline
        \end{tabular}}
    \end{center}
\end{table*}

\subsubsection{EEGDenoiseNet}
The dataset contains a total of 4514 clean EEG segments as ground truth, 3400 pure electrooculography (EOG) segments and 5598 pure electromyography (EMG) segments as ocular artifacts and myogenic artifacts respectively. The  contaminated signals are generated by linearly mixing the pure EEG segments with EOG or EMG artifact segments according to different signal-to-noise ratio (SNR) levels:
\begin{equation}
\label{eq:mixsnr}
    EEG_C= EEG_P + \lambda \cdot A,
\end{equation}
where $A$ refers to the artifacts, such as EMG and EOG; $\lambda$ is a hyperparameter to control the SNR in the contaminated EEG signal. To augment the size of dataset, The SNR of EEG contaminated by artifacts ranges from -7 to 2 dB, with 80\% of the dataset for generating the training set, 10\% for generating the validation set, and 10\% for generating the test set.

\subsubsection{Semi-Simulated EEG/EOG}
The dataset was constructed to evaluate the performance of the model in removing artifacts from multi-channel EEG. The dataset contains 19-channel EEG signals recorded from twenty-seven subjects and the corresponding EOG signal. The sampling rate of the signals is 200 Hz, and the signals are all segmented into one dimension segments of 30s. A band pass filtered at 0.5-40 Hz and notch filtered at 50 Hz are applied for EEG signals, while the EOG signals are band-pass filtered at 0.5-5 Hz. The contaminated EEG signals are produced by linearly combination of EOG signal and EEG signal according to the contamination method proposed by  \citep{elbert:1985:eog_denoise}. In this paper, the EEG signals are corrupted by EOG signals in the central 10s length for a 30s segment, which is marked as an artifact.

\subsubsection{MNE M/EEG dataset}
The MNE software package offers a sample dataset, which comprises recordings of combined MEG, EEG and EOG from a single participant, in which EEG signals were concurrently collected via 60 electrodes. This dataset is used to find a way to discriminate event-recorded potentials, including auditory stimuli (delivered monaurally to the left or right ear) and visual stimuli (shown in the left or right visual hemifield), which is used to validate the benefit of our method in practical uses.

\subsection{Performance Metrics}
The performance of our method was examined in two aspects: the efficiency of artifacts removal and application in downstream BCI classification task.
\subsubsection{Efficiency of Artifacts Removal}
First, Signal-to-noise ratio (SNR) is a measure that compares the level of a desired signal to the level of background noise, and the larger value is expected. The metric $\triangle{\text{SNR}}$ is defined to measure the improvement of SNR after artifact removal. The $\text{SNR}$ is calculated in decibels (dB) as follows:
\begin{equation}
    SNR = \frac{1}{n}\sum_{i=1}^n 10 \times \log_{10} \left ( {\frac{{\left \Vert \hat{x}_i \right \Vert}_2^2}{{\left \Vert \hat{x}_i - x_g \right \Vert}_2^2}} \right ),
\end{equation}
where $x_g$ is regarded as the ground truth signal and $n$ denotes the number of sampling points in each segment.

\begin{figure*}
    \centering
    \includegraphics[width=1.0\textwidth]{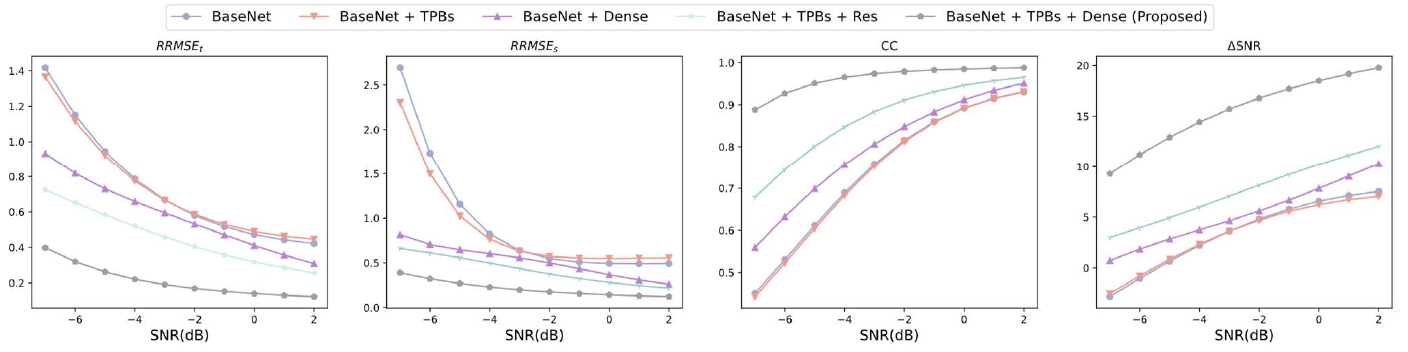}
    \centering
    \caption{Artifact removal performance at different SNR levels in ablation study.}
    \label{fig:ablation}
\end{figure*}

Additionally, we computed the relative root mean square error ($RRMSE$) between the generated signal and the ground truth in both the temporal ($RRMSE_t$, see eq.6) and spectral domains ($RRMSE_s$, see eq.7). Smaller values indicate the better de-noising performance.
\begin{equation}
    RRMSE_t = \frac{RMS(x_g - \hat{x})}{RMS(x)},
\end{equation}
\begin{equation}
    RRMSE_s = \frac{RMS(PSD(x_g) - PSD(\hat{x}))}{RMS(PSD(x))},
\end{equation}
where PSD is the power spectral density of the input signal.

Additionally, the impact of these methods on the nonlinear features of the reconstructed EEG signals is evaluated using the average correlation coefficient (CC, as defined in Eq. 8).
\begin{equation}
    CC = \frac{Conv(x_g, \hat{x})}{\sqrt{Var(x_g)Var(\hat{x})}},
\end{equation}
where functions $Conv$ and $Var$ denote the covariance and variance of the signal, respectively.
\subsubsection{Application in Downstream BCI Classification Task}To demonstrate the feasibility of \dtpabb, we further evaluated it on the BCI classification task. Here, the data pre-processed with the artifact removal method was used to classify the event-related potential (ERP). We used precision, recall , f1 scores, Matthews Correlation Coefficient (MCC) and Kappa statistics to evaluate the classification performance and the overall performance was evaluated by accuracy. The formulas for these evaluation indexes are as follows:
\begin{equation}
    Recall = \frac{TP}{TP+FN},
\end{equation}
\begin{equation}
    Precision = \frac{TP}{TP+FP},
\end{equation}
\begin{equation}
    F1 = \frac{2 \times Recall \times Precision}{Precision + Recall},
\end{equation}
\begin{equation}
    MCC = \frac{TP \times TN - FP \times FN}{\sqrt{(TP+FP)(TP+FN)(TN+FP)(TN+FN)}},
\end{equation}
\begin{equation}
    Kappa = \frac{2 \times (TP \times TN - FN \times FP)}{(TP+FP) \times (FP+TN)+(TP+FN)\times(FN+TN)},
\end{equation}
\begin{equation}
    Accuracy = \frac{TP + TN}{TP + FP + TN + FN},
\end{equation}
where $TP$ is true positive, $TN$ is true negative, $FP$ is false positive and $FN$ is false negative.

\subsection{Implementation Details}
All experiments were implemented using an open-source Python library, namely xai-kit \citep{xaikit}, on a server equipped with two NVIDIA GTX 2080Ti GPU cards. 
We employ MSE as the loss function. We adopt the Adam optimizer with the moments of $\beta_1=0.9$, $\beta_2=0.999$. The initial learning rate is set to 4.5e-4, and the batch size is set to 32. 
The early-stop patience of model training is set to 30. Due to the varying sampling rates and time-frequency characteristics of different artifacts across different datasets in the experiment, it’s necessary to fine-tune the network architecture parameters for different datasets and artifacts to adapt to their unique features. 
The key parameters to adjust include the Encoder filter number $N$, the Encoder filter length $L$, the channel number $H$ in TPB, the kernel size $P$ in TPB, the number of TPB $M$ for each repeat and the repeated times $R$. The parameters for different datasets are listed in \tabref{tab:hp}.

\subsection{Ablation Study}
In our network, the densely connected structure and the varying dilated rates in temporal pyramids are two crucial elements for better artifact removal of EEG. In this section, ablation experiments were conducted to analyze the effect of the above two components in our proposed method. Experiments were conducted with Dataset $\mathrm{I}$ contaminated by EMG artifacts, including the following group of configurations:

a) BaseNet: The unchanged Encoder and Decoder structure, combined with a standard feed-forward network as the artifact separator without shortcuts and temporal pyramids (dilation rate keeps at 1), used as the baseline for comparison.

b) BaseNet\,+\,TPBs: The BaseNet with temporal pyramids contained in the artifact separator.

c) BaseNet\,+\,Dense: The BaseNet with densely connected structure contained in the artifact separator.

d) BaseNet\,+\,TPBs\,+ Dense (Proposed): The BaseNet with densely connected TPBs contained in the artifact separator.

e) BaseNet\,+\,TPBs\,+\,Res: The BaseNet with TPBs connected in a ResNet fashion contained in the artifact separator.

Note that the densely connected structure was replaced by residual learning connection in e) to validate the superiority of the proposed structure to the ResNet structure.

The quantitative result of EEG artifact removal at multiple SNR levels (-7 to 2 dB) was shown in \figref{fig:ablation}. It could be observed that without the densely connected structure, the BaseNet would have similar denoising performance regardless of whether Temporal Pyramids are incorporated. This phenomenon could be attributed to the deep network structure (total of 109 convolutional layers), which makes it difficult for the shallow layers to learn effective features. Therefore, even if the forward network possesses features of different scales, due to the limited learning ability, it is difficult to achieve good performance.  Compared to the above two networks, BaseNet+Dense allows additional supervision from the loss function with shorter connections which enforce the intermediate layers to learn discriminative features, improving performance effectively. Additionally, the synergy between Temporal Pyramids and densely connected network (BaseNet+TPBs+Dense) resulted in a significant performance improvement, compared to the results of BaseNet+TPBs and BaseNet+Dense, demonstrating the effectiveness of combination of the two components.

It could also be observed that the denoising performance of the network deteriorates when the densely connected structure was replaced by residual links (BaseNet+TPBs+Res), which indicates that the densely connected network is crucial in our network. The densely connected network and Temporal Pyramids synergistically promote the learning of effective features in EEG signals, and their roles in the network are dispensable.

\subsection{Performance Evaluation}
To validate the effectiveness of the \dtpabb \ method, we benchmarked it against alternative artifact removal algorithms which are designed for denoising EEG recordings: SimpleCNN \citep{zhang:2021:eegdenoisenet}, 1D-ResCNN \citep{sun:2020:dleegdenoise}, NovelCNN \citep{zhang:2021:cnndenoise} and WQN \citep{dora:2022:adaptive_eegdenoise}. For a fair comparison, the most recent state-of-the-art approaches whose implementation code is available online are selected. The model complexities of different baseline networks is reported in \tabref{tab:tp}.

\subsubsection{Model deployment I via EEGDenoiseNet Dataset}
\begin{figure*}
    \centering
    \includegraphics[width=\textwidth]{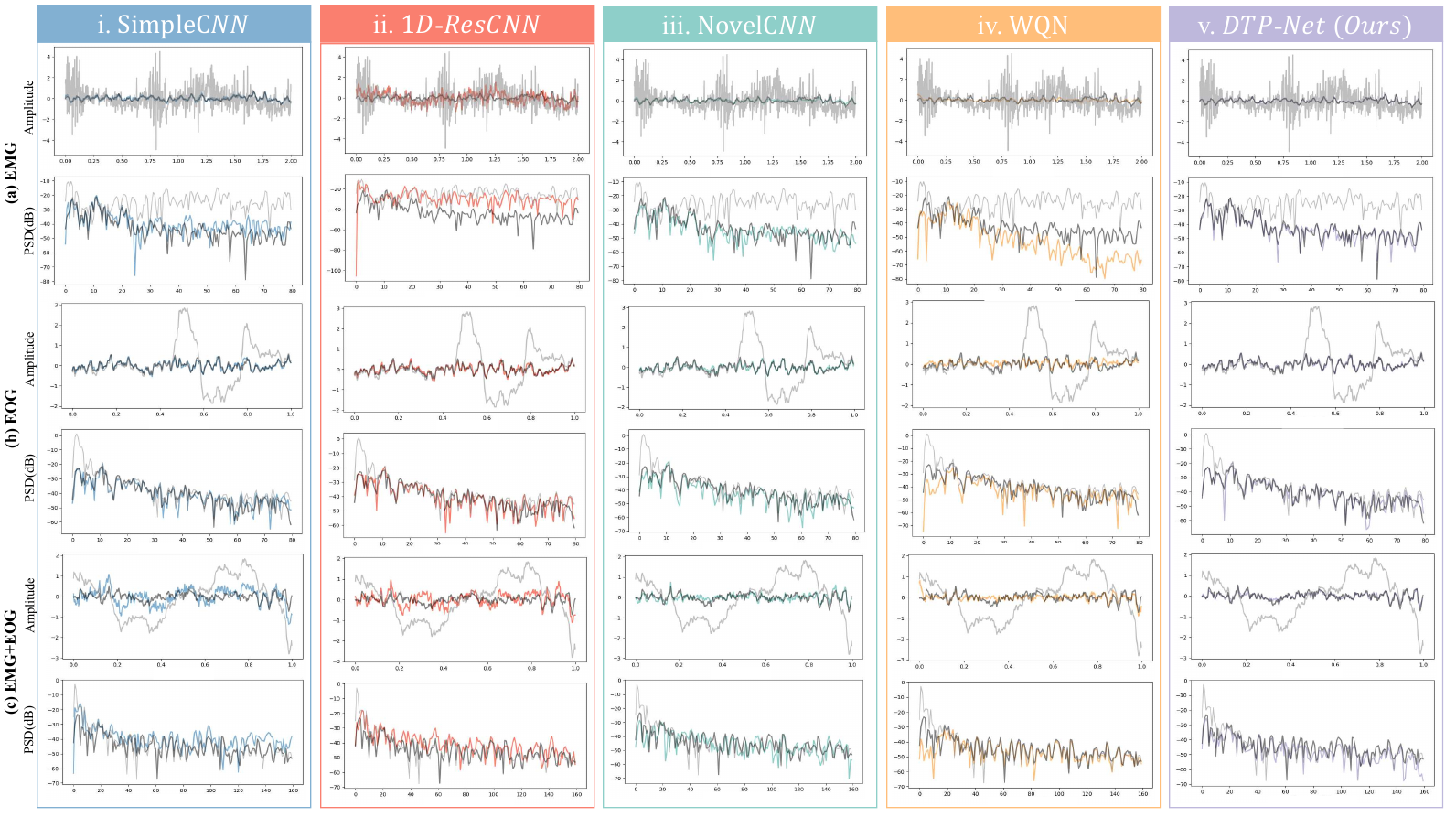}
    \centering
    \caption{Example of experimental waveform results and PSD results for eliminating artifacts from EEGDenoiseNet Dataset.}
    \label{eeg_denoise_v}
\end{figure*}

\begin{figure*}
    \centering
    \includegraphics[width=\textwidth]{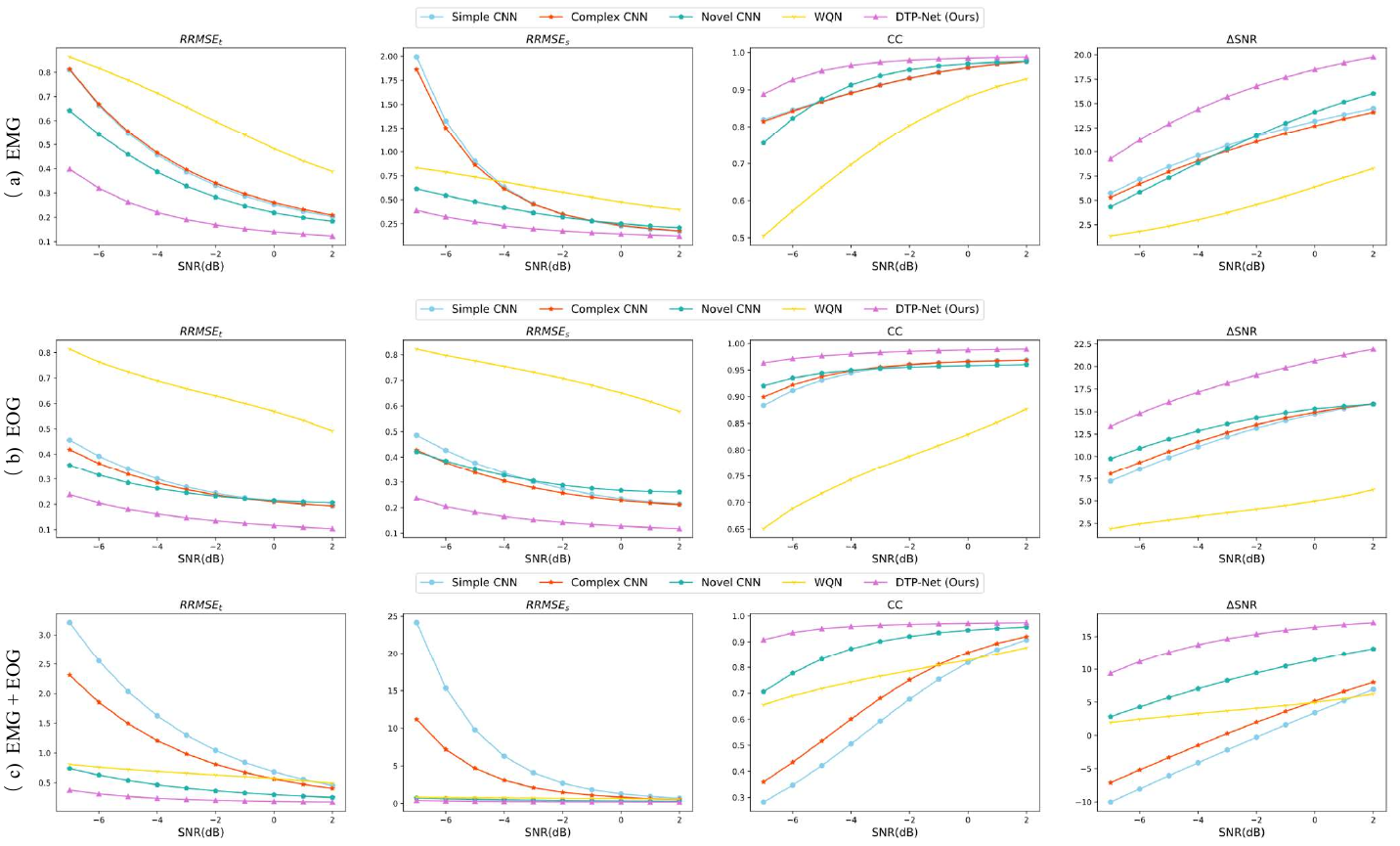}
    \centering
    \caption{Experimental results at different SNR levels after artifact removal from EEGDenoiseNet Dataset.}
    \label{eeg_denoise_c}
\end{figure*}

\setlength{\arrayrulewidth}{0.4mm}
\renewcommand{\arraystretch}{1.7}
\begin{table*}
\caption{Average Performance of Multiple Artifacts at Multiple SNR Levels}
\label{tab:metrics}
    \begin{center}
    \resizebox{\linewidth}{!}{
        \begin{tabular}{p{1.5cm}<{\centering} c c c c c c c}
        \toprule[2pt]
        \textbf{Dataset} & \textbf{Artifact} & \textbf{Metric} & \textbf{1D-ResCNN} & \textbf{SimpleCNN} & \textbf{NovelCNN} & \textbf{WQN} & \textbf{\dtpabb \ (Ours)} \\
        \hline
        \multirow{12}*{\makecell[c]{\textbf{EEG} \\ \textbf{DenoiseNet}}} & 
        \multirow{4}*{EMG} & $RRMSE_t$ & $0.4544 \pm 0.0159$ & $0.4229 \pm 0.0028$ & $0.3424 \pm 0.0026$ & $0.6262$ & \bm{$0.2106 \pm                              0.0125$} \\ 
        {} & {} & $RRMSE_s$ & $0.6914 \pm 0.0367$ & $0.6165 \pm 0.0162$ & $0.3682 \pm 0.0070$ & $0.6083$ & \bm{$0.2136 \pm 0.0219$} \\ 
        {} & {} & $\triangle{SNR}$ & $10.1323 \pm 0.2327$ & $10.7494 \pm 0.0485$ & $10.8190 \pm 0.1274$ & $4.4242$ & \bm{$15.5477 \pm                0.1779$} \\ 
        {} & {} & $CC$ & $0.9025 \pm 0.2327$ & $0.9101 \pm 0.0012$ & $0.9164 \pm 0.0012$ & $0.7533$ & \bm{$0.9626 \pm 0.0070$} \\ \cline{2-8}
        {} & \multirow{4}*{EOG} & $RRMSE_t$ & $0.2583 \pm 0.0065$ & $0.2849 \pm 0.0009$ & $0.2468 \pm 0.0040$ & $0.6473$ & \bm{$0.1522 \pm                           0.0115$} \\
        {} & {} & $RRMSE_s$ & $0.2784 \pm 0.0027$ & $0.3139 \pm 0.0017$ & $0.2993 \pm 0.0011$ & $0.7117$ & \bm{$0.1585 \pm 0.0137$} \\ 
        {} & {} & $\triangle{SNR}$ & $12.9844 \pm 0.1933$ & $12.1151 \pm 0.0463$ & $13.9070 \pm 0.2374$ & $3.9573$ & \bm{$18.2231 \pm                0.1021$} \\ 
        {} & {} & $CC$ & $0.9542 \pm 0.0027$ & $0.9450 \pm 0.0004$ & $0.9566 \pm 0.0015$ & $0.7723$ & \bm{$0.9812 \pm 0.0050$} \\ \cline{2-8}
        {} & \multirow{4}*{EMG + EOG} & $RRMSE_t$ & $1.0802 \pm 0.0229$ & $1.4149 \pm 0.0260$ & $0.4219 \pm 0.0042$ & $0.6457$ &                                                \bm{$0.2342 \pm 0.0067$} \\
        {} & {} & $RRMSE_s$ & $3.4649 \pm 0.2276$ & $6.6571 \pm 0.2764$ & $0.4453 \pm 0.0042$ & $0.7085$ & \bm{$0.2519 \pm 0.0039$} \\ 
        {} & {} & $\triangle{SNR}$ & $0.8468 \pm 0.1673$ & $1.2657 \pm 0.1464$ & $8.5953 \pm 0.0719$ & $3.9681$ & \bm{$14.3121 \pm                   0.3888$} \\ 
        {} & {} & $CC$ & $0.6853 \pm 0.0071$ & $0.6188 \pm 0.0028$ & $0.8842 \pm 0.0030$ & $0.7728$ & \bm{$0.9812 \pm 0.0050$} \\
        \hline
        \multirow{4}*{\makecell[c]{\textbf{Semi-} \\ \textbf{Simulated} \\ \textbf{EOG}}} &
        \multirow{4}*{EOG} & $RRMSE_t$ & $0.2999 \pm 0.0414$ & $0.2703 \pm 0.0145$ & $0.1676 \pm 0.0128$ & $0.2650$ & \bm{$0.0563 \pm                           0.0064$} \\ 
        {} & {} & $RRMSE_s$ & $0.7848 \pm 0.1476$ & $0.7191 \pm 0.0253$ & $0.1848 \pm 0.0188$ & $0.5130$ & \bm{$0.0747 \pm 0.0086$} \\
        {} & {} & $\triangle{SNR}$ & $12.2579 \pm 1.2187$ & $13.1235 \pm 0.4336$ & $13.6394 \pm 0.7394$ & $12.7315$ & \bm{$29.3461 \pm                 1.6250$}\\ 
        {} & {} & $CC$ & $0.9459 \pm 0.0154$ & $0.9549 \pm 0.0030$ & $0.9802 \pm 0.0020$ & $0.9560$ & \bm{$0.9949 \pm 0.0003$}\\
        \bottomrule[2pt]
        \end{tabular}}
    \end{center}
\end{table*}

\begin{figure*}
    \centering
    \includegraphics[width=0.95\textwidth]{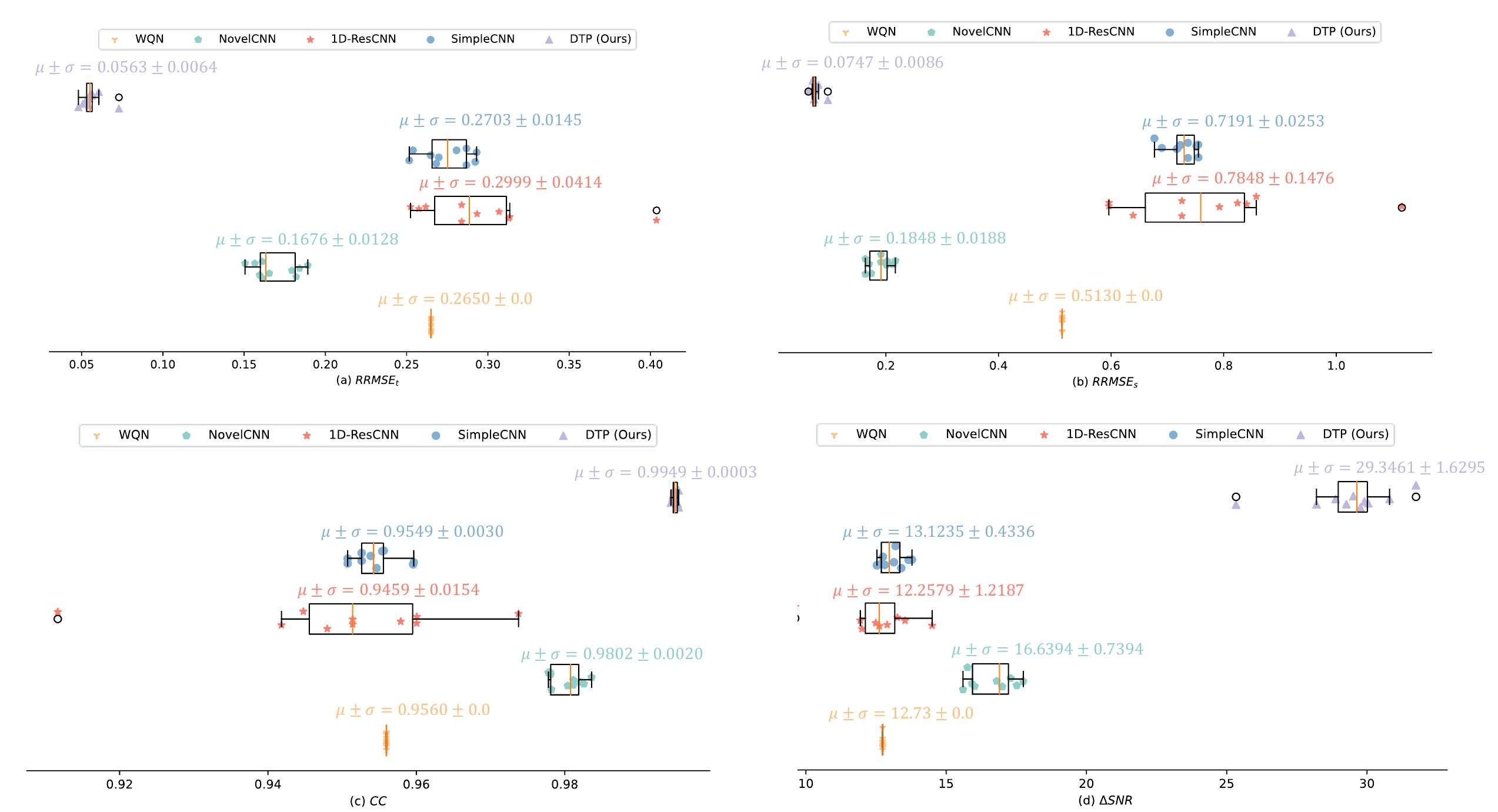}
    \centering
    \caption{Box-and-scatter plot of $RRMSE_t$, $RRMSE_s$, $CC$ and $\triangle{SNR}$ on EOG artifact removal from Semi-simulated Dataset. Among the five models, \dtpabb \ can achieve better and more robust performance.}
    \label{semi_simulated_scatter}
\end{figure*}

\begin{figure*}
    \centering
    \includegraphics[width=\textwidth]{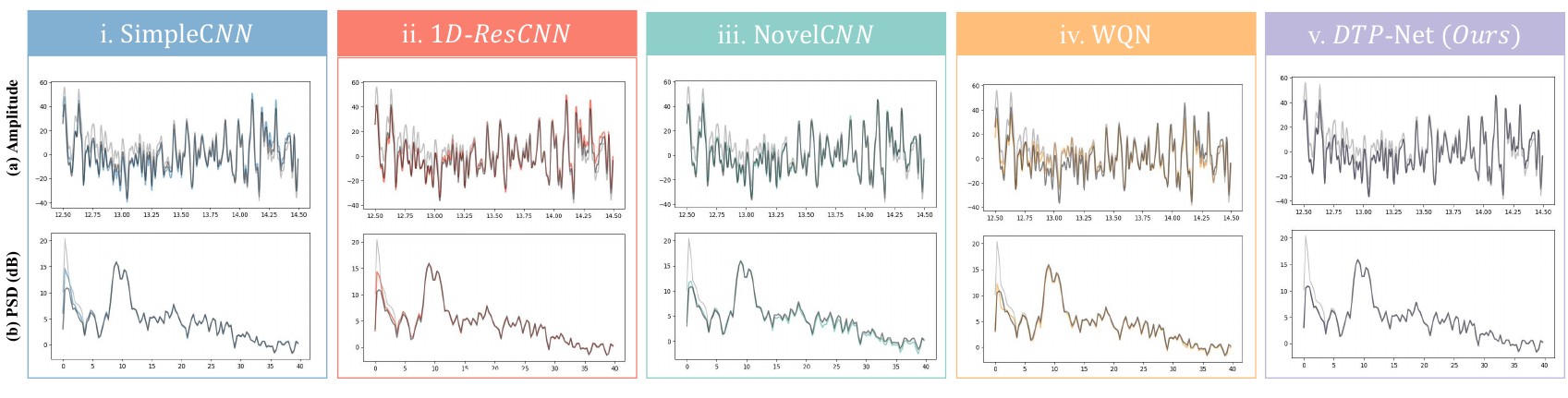}
    \centering
    \caption{Example of the waveform results and PSD results on Pz channel after eliminating ocular artifacts from multi-channel contaminated EEG signals.}
    \label{semi_simulated_v}
\end{figure*}

To provide a qualitative analysis, the reconstruction results were shown in \figref{eeg_denoise_v} for the 2-s EEG time series respectively contaminated by EMG, EOG and EMG+EOG noises in $SNR=-7dB$ condition. To provide a comprehensive analysis, the experimental results were presented in the temporal domain and frequency domain. The light gray line indicated the raw EEG with different kinds of artifacts, while the black line indicated the ground-truth clean EEG signal. The different colored lines indicated the reconstructed signal using SimpleCNN, 1D-ResCNN, NovelCNN, WQN and the proposed \dtpabb \ method respectively.

From temporal results in \figref{eeg_denoise_v}, it could be observed that the removal of artifacts denosing by SimpleCNN and 1D-ResCNN is incomplete and the corrected EEG signals are still severely distorted compared to the pure EEG signal, especially for elimination of EMG and EOG compound noises (\figref{eeg_denoise_v} (c)).  Although the other methods have similar performance with \dtpabb \ method, the corrected waveform is floating compared to the original EEG segments, and \dtpabb \ performs better in reconstructing the signal details. Especially from the local details of the denosied EEG signal, it can be found that \dtpabb \ have the best fit of the colored and black line, indicating the least loss in the temporal domain. In the spectrum analysis view, the PSD lines of \dtpabb \ had the highest degree of fitting with the black line in three artifacts sources, which indicated that the reconstruction of EEG signal using \dtpabb \ effectively removed the noise components and well observed the useful frequency information.

The SNR metric denotes the level at which the pure EEG signals are contaminated by artifacts. To further investigate the robustness of our model in noisy EEG, we presented the quantitative metrics ($RRMSE_{t}$, $RRMSE_{s}$, $CC$ and $\triangle SNR$) from four comparative methods and our proposed \dtpabb \ at multiple SNR levels (-7 to 2 dB). The quantitative results were shown in \figref{eeg_denoise_c}, including the artifacts removal for the EEG signals contaminated by EMG, EOG and EMG-EOG compound noises respectively. It could be observed that the performance of all five methods became worse with decreasing SNR for all types of artifacts. Some methods behaved well at low SNR levels, while having inferior capacity at high SNR levels (such as NovelCNN method in EOG artifacts elimination task). But \dtpabb \ method reliably outperformed other methods across different SNR levels and noise types, showing lower $RRMSE_t$, lower $RRMSE_s$, higher $CC$ and higher $\triangle SNR$. The result suggested \dtpabb \ is robust even when there are large artifacts in the input signal.

To get the quantitative results, the metric values for all methods across all SNRs (-7 dB to 2dB) were averaged, which were summarized in \tabref{tab:metrics}. To test the robustness of model, the four networks were trained independently for 10 times with randomly generated dataset via EEGDenoiseNet and performance was tested in a fixed dataset with different SNR levels. The average performance and variance of the models were recorded in \tabref{tab:metrics}. For WQN, a traditional method, since it did not have the uncertainty caused by parameter initialization and training process, only the average performance was recorded in \tabref{tab:metrics}. The results in \tabref{tab:metrics} showed that our proposed model was much better than other competing models in terms of all four metrics and three types of artifacts,  obtaining the lowest $RRMSE_{t}$, lowest $RRMSE_{s}$, highest $CC$ and highest $\triangle SNR$. It showed that \dtpabb \ could substantially suppress the distortion in the temporal and spectral waveform, and retain the clean EEG signal as much as possible. This result was consistent with the visualization result in \figref{eeg_denoise_v}.

\subsubsection{Model deployment II via Semi-simulated Dataset}
The experiment in this section was to investigate the effectiveness of the model in multi-channel EEG denoising. The performance of experiments on the Semi-Simulated EEG/EOG dataset was reported in \tabref{tab:metrics}, in which the comprehensive results were obtained by averaging the results from different spatial channels and different patients. The quantitative results showed that \dtpabb \ still obtained superior performance than other methods, no matter in the temporal domain or the spectral domain. The different methods were repeated 10 times to obtain the average performance of the models and the distribution results of the repeated experiments were shown in \figref{semi_simulated_scatter}. Since the WQN method was a deterministic process which would not be disturbed by random factors, its denoising results for the same data were constant and its standard deviation for repeated experiments was zero. It could be observed that \dtpabb \ outperformed other baseline networks in terms of $CC$, $\triangle{SNR}$, $RRMSE_t$ and $RRMSE_s$ which confirmed the efficiency in information restoration and SNR improvement. In addition, the distribution of \dtpabb \ under four metrics was more concentrated than other methods, which indicated that \dtpabb \ is more robust in multi-channel EEG artifacts removal.

\renewcommand{\arraystretch}{1.5}
\begin{table*}
\caption{Classification Performance of BCI Dataset}
\label{tab:bci}
    \begin{center}
        \begin{tabular}{c c c c c c c}
        \toprule[2pt]
        \multirow{2}*{} & \textbf{Artifact Removal Method} & \textbf{Raw} & \textbf{1D-ResCNN} & \textbf{SimpleCNN} & \textbf{NovelCNN} & \textbf{\dtpabb \ (Ours)} \\ \cline{2-7}
        {} & \textbf{Classifier Method} & \multicolumn{5}{c}{\textbf{pyRiemann}} \\
        \midrule[2pt]
        \multirow{6}*{\makecell[c]{\textbf{Overall} \\ \textbf{Metrics}}} & Accuracy & 0.7778 & 0.6944 & 0.7222 & 0.7639 & \textbf{0.8333} \\
        {} & Recall & 0.7978 & 0.7270 & 0.7498 & 0.8003 & \textbf{0.8455} \\
        {} & Precision & 0.7993 & 0.7341 & 0.7544 & 0.8181 & \textbf{0.8491} \\
        {} & F1 Score & 0.7985 & 0.7305 & 0.7521 & 0.8090 & \textbf{0.8473} \\
        {} & MCC & 0.7081 & 0.5929 & 0.6383 & 0.6885 & \textbf{0.7828} \\
        {} & Kappa & 0.7039 & 0.5389 & 0.6306 & 0.6791 & \textbf{0.7782} \\
        \hline
        \multirow{4}*{\makecell[c]{\textbf{Per-class} \\ \textbf{F1-score}}} & Audio Left & 0.6739 & 0.6597 & 0.6478 & \textbf{0.7807} & 0.7507 \\
        {} & Audio Right & 0.7205 & 0.6728 & 0.6260 & 0.7972 & \textbf{0.8117} \\
        {} & Visual Left & 0.8906 & 0.8246 & 0.8384 & 0.8694 & \textbf{0.9006} \\
        {} & Visual Right & 0.8908 & 0.7312 & 0.8655 & 0.7464 & \textbf{0.9134} \\
        \bottomrule[2pt]
        \end{tabular}
    \end{center}
\end{table*}

The denoising performance on multiple channels EEG data for semi-simulated EEG/EOG dataset demonstrated that most of the ocular artifacts could be removed automatically through \dtpabb, confirming that \dtpabb \ could maintain stable and efficient denoising ability on different spatial channels. To further investigate the denoising effect, the contaminated EEG channel (Pz) and the denoising results of different methods were depicted in \figref{semi_simulated_v}. As can be seen in \figref{semi_simulated_v}, various denoise methods behaved similarly by the visual inspection of temporal waveforms. In the spectrum analysis view, it showed that the frequency distortion caused by EOG noise was mainly concentrated in the low-frequency band ($\delta$ band). SimpleCNN and 1D-ResCNN eliminated EOG noise partially, and the removal of spectral distortion was incomplete. NovelCNN and WQN reduced most of the noise, but the denoised PSD lines had a certain up-and-down float compared to the original EEG. In contrast, the EEG spectrum after the proposed model denoising maintained the smallest deviation and the highest data quality.

\subsubsection{Model deployment III via BCI experiment}
The effect of EEG artifact elimination might influence the downstream analysis, such as brain-computer interface (BCI) performance. Here we aim to investigate the beneficial effects of EEG denoising on event-related potential (ERP), which is a stereotyped neural response to novel stimuli. In this experiment, pyRiemann  \citep{alexandre:2022:pyriemann} was used to classify ERP EEG data from a four-class classification task, which is a software package for processing and classification of multivariate time series through the Riemannian geometry of symmetric positive definite (SPD) matrices. Prior to the classification of ERP, various denoising methods were applied to preprocess the EEG signals, and the performance of different methods on downstream tasks was compared. Due to the unavailability of clean EEG data for model training in this task, the denoising model trained on Semi-simulated dataset would be directly employed, which could also serve as a test for the model’s generalization ability. As the WQN method required temporally adjacent uncontaminated signal as reference which was not available in this task, it was not used as a comparative method in this experiment. 

As shown in \tabref{tab:bci}, the classification method with Novel CNN and \dtpabb \ pre-processed data achieved better performance than raw signal with higher recall and precision, implying the effective artifacts removal and signal feature enhancement. In contrast, the data denoised by 1D-ResCNN and SimpleCNN degraded the performance in classification compared to the raw signal, indicating that these methods not only removed artifacts but also trimmed the EEG signal information. The classification results showed that \dtpabb \ significantly outperformed other methods in downstream analysis, achieved the F1 score improvement from 0.7985 (recall: 0.7978; precision: 0.7993) in raw signal to 0.8473 (recall: 0.8455; precision: 0.8491).

\section{Discussion}

\begin{figure}
    \centering
    \includegraphics[scale=0.37]{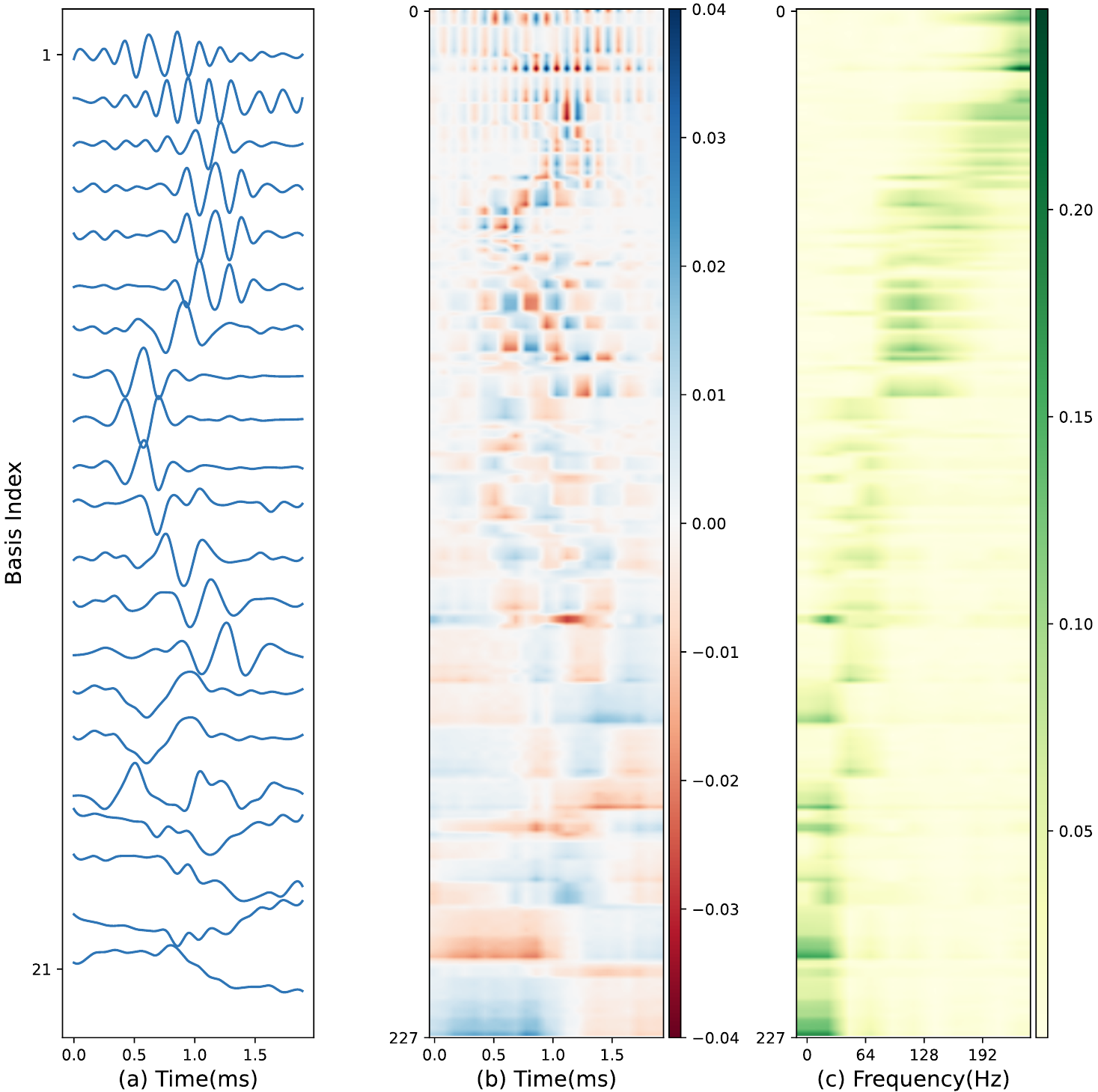}
    \centering
    \caption{Visualization of filter bank in Encoder, sorted by frequency. (a) Examples of basis functions' waveform. (b) Visualization of Encoder basis functions. (c) Visualization of Encoder basis functions' spectrums.}
    \label{Encoder_weights}
\end{figure}

The proposed \dtpabb \ is an end-to-end artifact removal method that is capable of directly mapping contaminated EEG signals to clean EEG. By conducting experiments on different datasets, we demonstrated the  consistently effective performance of our proposed method in removing artifacts and recovering clean signals for both single-channel and multi-channel signals, in various noise sources and SNR levels. From the ablation study, it could be observed that each component of the \dtpabb \ model plays an indispensable role in noise reduction. In this section, we aim to understand the underlying mechanisms and dynamics of each module in the network, thereby enhancing model reliability and providing guidance for the future design of sequence-based task models. The analysis of network behavior presented below is based on the experiments conducted on Dataset I contaminated by EMG artifacts.
\subsection{Model behavior of Encoder layer}
As mentioned in Section II.B, the Encoder module within the network is specifically designed to extract time-frequency features through convolutional operations. The layer’s weights can be considered as a filter bank comprising distinct frequency components, and these components are learned through joint training with subsequent modules of the network.  

To facilitate understanding, the sorted Encoder weights are visualized to probe the behavior of Encoder layer in terms of time-frequency feature extraction, as shown in \figref{Encoder_weights}. It could be observed that an effective subset of basis function is learned in Encoder layer. The kernels span a wide range of frequencies to capture the diverse spectral characteristics (shown in \figref{Encoder_weights}(c)). Additionally, the temporal transition inside the kernel window (shown in \figref{Encoder_weights}(a)) can be adaptively achieved during the training process to zoom in signal details. This property can provide a tilling of time-frequency space that offers a better compromise between time and frequency resolution to well adapted to the characteristics of the non-stationary characteristics of EEG signals. Additionally, as shown in \figref{eeg_denoise_v}, despite employing the temporal MSE as loss function, the EEG spectrum after \dtpabb \ denoising is the closest to the clean EEG spectrum compared to other methods. This can be attributed to the precise extraction of the time-frequency characteristics by Encoder layer.

\begin{figure}
    \centering
    \includegraphics[scale=0.56]{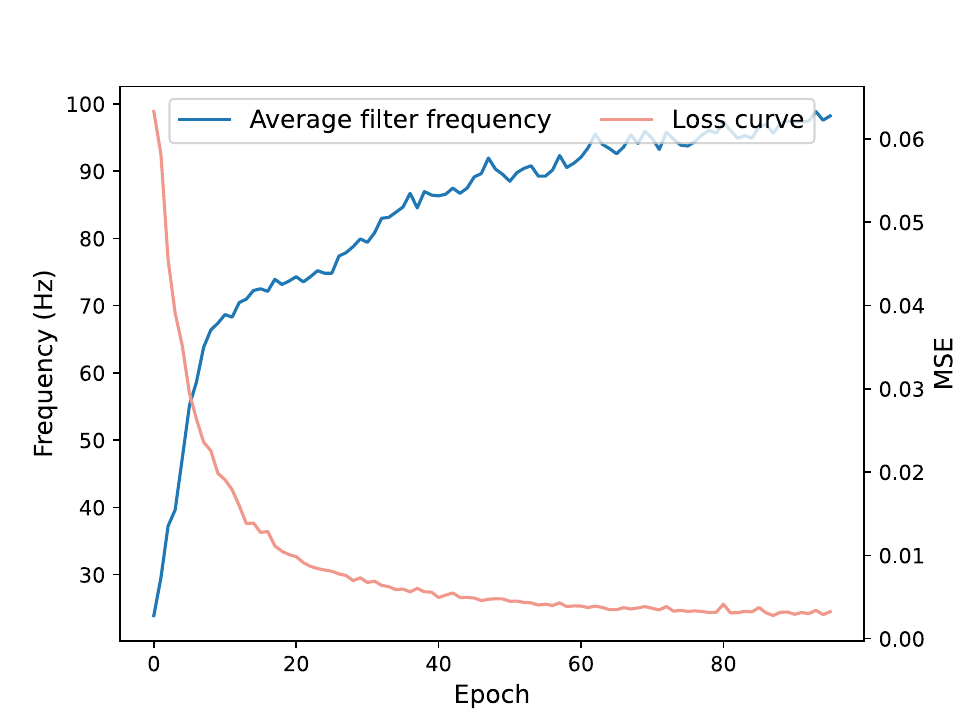}
    \centering
    \caption{The average filter frequency increases as the validation loss decreases during training,
    which empirically coincides with the Frequency Principle.    }
    \label{filter frequency}
\end{figure}

Moreover, we investigate the evolution process of the learned frequency components in the Encoder layer during the training phase, as depicted in \figref{filter frequency}. It could be observed that, during the initial stages of training, the layer demonstrates a propensity for fitting low-frequency components, gradually transitioning to the fitting of high-frequency components as the training proceeds. It provides an empirical evidence to an implicit bias in the gradient-based training of DNNs, that is, Frequency Principle \citep{xu:2019:train_behaviour}: \textit{DNNs often fit target functions from low to high frequencies during the training process.}

\subsection{Model behavior of Temporal Pyramids}
The synergistic effect between Temporal Pyramids with periodical exponentially increased dilation blocks and densely connected architecture has been found in ablation experiment in III.D. Here, we visualize and probe the contribution of Temporal Pyramids to the cooperated artifacts removal from the perspective of representation learning. 

\begin{figure}
    \centering
    \includegraphics[scale=0.56]{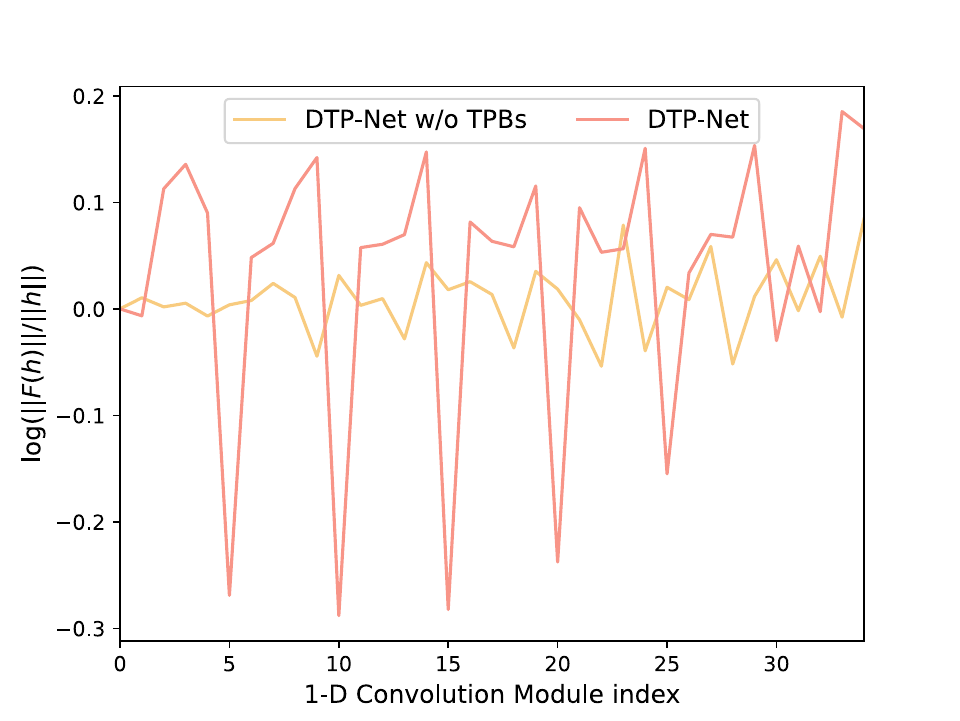}
    \centering
    \caption{The comparison of representation learning ability between \dtpabb \ and \dtpabb \ w/o TPBs via average ratio of $\ell^2$ norm of output of 1-D convolution module to the norm of input of 1-D convolution module.}
    \label{tpb blocks}
\end{figure}

For the $i$-th TPB $F_i(\cdot)$ in Temporal Pyramids which transforms representation as $h_{i+1} = F_i(h_i)$, the representation learning performance (RLP) is evaluated via logarithm to $\ell^2$ ratio of $||F_i(h_i)||_2/||h_i||_2$  \citep{jastrz:2018:residual_connection} averaged across samples. This ratio serves as a direct indicator of the extent to which $F_i(\cdot)$ changes the representation $h_i$, and a substantial change can be argued to be a prerequisite for the block to effectively engage in representation learning. An RLP comparison between our \dtpabb \ model with and without Temporal Pyramids (dilation rate keeps at 1 for dilated convolution in each TPB) is presented in \figref{tpb blocks}, illustrating the evolution of the performance across convolution blocks as we go to higher blocks. It could be observed that our \dtpabb \ exhibits an explicitly periodic pattern of feature learning with the repetition of Temporal Pyramids, while each convolutional block performs moderate learning ability if there’s no Temporal Pyramid contained in \dtpabb.

We intend to analyze the above learning behavior from frequency perspective.  From section II. D, it could be referred that $d$-dilated convolution is equivalent to signal downsampled by a factor of $d$, effectively shifting the learning focus from high-frequency components to low-frequency ones. The F-Principle \citep{xu:2019:train_behaviour} states that DNNs excel at processing low-frequency content compared with high-frequency ones, exhibiting rapid learning and good generalization error. Thus, the Temporal Pyramids with TPBs possessing exponentially increased dilation rates perform periodic hierarchical downsampling on the input signals, consequently showcasing the periodic hierarchical representation learning ability shown in \figref{tpb blocks}. For \dtpabb \ without Temporal Pyramids, high-frequency components are preserved without signal downsample, thus RLP leading to the consistent moderate feature learning capacity.

In conclusion, the Temporal Pyramids facilitate expeditious acquisition of deep abstract features and amplify their generalization prowess. Hence, the corporation between Temporal Pyramids and densely connected structure shown in ablation experiment could be analyzed explicitly. The Temporal Pyramids make it easier for hidden blocks to learn from complex feature patterns, while the densely connected structure makes the learning process transparent by enforcing additional direct supervision via dense shortcuts.
\subsection{Limitations}
Although the proposed model exhibits feasibility in eliminating diverse artifacts, there are several limitations to be improved in our future work. 

For example, for multi-channel signals, the complex spatial information of artifacts and EEG signals are distinct characteristics in each channel, while \dtpabb \ processes each EEG channel separately without concerning correlation amongst channels. Thus, how to incorporate spectral-temporal-spatial information in EEG denoising tasks should be further examined in future research. Additionally, our proposed \dtpabb \ EEG denoising approach is expected to be applied broadly in the fields of neuroscience and neuro-engineering, while we only explore its application in BCI domain in this paper. Further investigations are required in future work to explore its potential use in medical fields such as sleep disorder diagnosis and sleep staging.

\section{Conclusion}
In this study, we propose a novel fully convolutional \dtpabb \ method which can be used to automatically reduce various deep artifacts from EEG data through feature-based noise separation method. \dtpabb \ uses Encoder-Decoder layer to achieve complex signal transformations between time domain and time-frequency domain and artifact separator block to extract diverse time-varying characteristics of artifacts. 

The results from experimental studies conducted on two semi-simulated EEG datasets consistently reveal that our proposed DTP surpasses state-of-the-art algorithms in effectively mitigating artifacts of diverse nature and varying SNR levels. Furthermore, an experiment of BCI application was conducted, demonstrating higher classification accuracy and thereby validating the robustness, generalizability, and potential applications of our \dtpabb \ method in the fields of EEG-based neuroscience and neuro-engineering.

Additionally, the underlying mechanisms and dynamics of DTP-Net are investigated. The distinguishable time-frequency features learned by Encoder layer and the periodically hierarchical representation learning process achieved by the artifact separator are observed. These mechanisms facilitate DTP-Net to efficiently address denoising tasks in nonstationary EEG signals. However, as a single-channel denoised model, DTP-Net ignores the complex spatial characteristics of artifacts and EEG signals. Applying the DTP-Net to reduce artifacts in multi-channel EEG signals, incorporating spectral-temporal-spatial information in EEG signals, should be further examined in future research. Overall, \dtpabb \ method is demonstrated to be an effective preprocessing way for EEG analysis to retrieve artifact-reduced EEG from contaminated raw EEG.
\bibliographystyle{unsrtnat}

\begin{thebibliography}{44}
\providecommand{\natexlab}[1]{#1}
\providecommand{\url}[1]{\texttt{#1}}
\expandafter\ifx\csname urlstyle\endcsname\relax
  \providecommand{\doi}[1]{doi: #1}\else
  \providecommand{\doi}{doi: \begingroup \urlstyle{rm}\Url}\fi

\bibitem[Chen et~al.(2022)Chen, Li, Liu, McKeown, Qian, and Wang]{chen:2022:eegdl}
Xun Chen, Chang Li, Aiping Liu, Martin~J. McKeown, Ruobing Qian, and Z.~Jane Wang.
\newblock Toward open-world electroencephalogram decoding via deep learning: A comprehensive survey.
\newblock \emph{{IEEE} Signal Processing Magazine}, 39\penalty0 (2):\penalty0 117--134, 2022.
\newblock \doi{10.1109/msp.2021.3134629}.
\newblock URL \url{https://doi.org/10.1109/msp.2021.3134629}.

\bibitem[Schomer and da~Silva(2017)]{donald:2017:eeg}
Donald~L. Schomer and Fernando H.~Lopes da~Silva, editors.
\newblock \emph{Niedermeyer{\textquotesingle}s Electroencephalography}.
\newblock Oxford University Press, November 2017.
\newblock \doi{10.1093/med/9780190228484.001.0001}.
\newblock URL \url{https://doi.org/10.1093/med/9780190228484.001.0001}.

\bibitem[Jiang et~al.(2019)Jiang, Bian, and Tian]{jiang:2019:denoise_review}
Xiao Jiang, Gui-Bin Bian, and Zean Tian.
\newblock Removal of artifacts from {EEG} signals: A review.
\newblock \emph{Sensors}, 19\penalty0 (5):\penalty0 987, February 2019.
\newblock \doi{10.3390/s19050987}.
\newblock URL \url{https://doi.org/10.3390/s19050987}.

\bibitem[Jung et~al.(1998)Jung, Makeig, Bell, and Sejnowski]{jung:1998:ica_eeg}
Tzyy-Ping Jung, Scott Makeig, Anthony~J. Bell, and Terrence~J. Sejnowski.
\newblock Independent component analysis of electroencephalographic and event-related potential data.
\newblock In \emph{Central Auditory Processing and Neural Modeling}, pages 189--197. Springer {US}, 1998.
\newblock \doi{10.1007/978-1-4615-5351-9_17}.
\newblock URL \url{https://doi.org/10.1007/978-1-4615-5351-9_17}.

\bibitem[Mannan et~al.(2018)Mannan, Kamran, and Jeong]{mannan:2018:denoise_review}
Malik Muhammad~Naeem Mannan, Muhammad~Ahmad Kamran, and Myung~Yung Jeong.
\newblock Identification and removal of physiological artifacts from electroencephalogram signals: A review.
\newblock \emph{{IEEE} Access}, 6:\penalty0 30630--30652, 2018.
\newblock \doi{10.1109/access.2018.2842082}.
\newblock URL \url{https://doi.org/10.1109/access.2018.2842082}.

\bibitem[Clercq et~al.(2006)Clercq, Vergult, Vanrumste, Paesschen, and Huffel]{wim:2006:cca_eeg}
Wim~De Clercq, A.~Vergult, B.~Vanrumste, W.~Van Paesschen, and S.~Van Huffel.
\newblock Canonical correlation analysis applied to remove muscle artifacts from the electroencephalogram.
\newblock \emph{{IEEE} Transactions on Biomedical Engineering}, 53\penalty0 (12):\penalty0 2583--2587, November 2006.
\newblock \doi{10.1109/tbme.2006.879459}.
\newblock URL \url{https://doi.org/10.1109/tbme.2006.879459}.

\bibitem[Mumtaz et~al.(2021)Mumtaz, Rasheed, and Irfan]{mumtaz:2021:eeg_denoise_review}
Wajid Mumtaz, Suleman Rasheed, and Alina Irfan.
\newblock Review of challenges associated with the {EEG} artifact removal methods.
\newblock \emph{Biomedical Signal Processing and Control}, 68:\penalty0 102741, July 2021.
\newblock \doi{10.1016/j.bspc.2021.102741}.
\newblock URL \url{https://doi.org/10.1016/j.bspc.2021.102741}.

\bibitem[Inuso et~al.(2007)Inuso, Foresta, Mammone, and Morabito]{Inuso2007}
Giuseppina Inuso, Fabio~La Foresta, Nadia Mammone, and Francesco~Carlo Morabito.
\newblock Wavelet-{ICA} methodology for efficient artifact removal from electroencephalographic recordings.
\newblock In \emph{2007 International Joint Conference on Neural Networks}. {IEEE}, August 2007.
\newblock \doi{10.1109/ijcnn.2007.4371184}.
\newblock URL \url{https://doi.org/10.1109/ijcnn.2007.4371184}.

\bibitem[Sweeney et~al.(2013)Sweeney, McLoone, and Ward]{sweeney:2013:emd-cca}
Kevin~T. Sweeney, Se{\'{a}}n~F. McLoone, and Tom{\'{a}}s~E. Ward.
\newblock The use of ensemble empirical mode decomposition with canonical correlation analysis as a novel artifact removal technique.
\newblock \emph{{IEEE} Transactions on Biomedical Engineering}, 60\penalty0 (1):\penalty0 97--105, January 2013.
\newblock \doi{10.1109/tbme.2012.2225427}.
\newblock URL \url{https://doi.org/10.1109/tbme.2012.2225427}.

\bibitem[Nguyen et~al.(2012)Nguyen, Musson, Li, Wang, Zhang, Xu, Richey, Schnell, McKenzie, and Li]{nguyen:2012:wavelet_neural_network}
Hoang-Anh~T. Nguyen, John Musson, Feng Li, Wei Wang, Guangfan Zhang, Roger Xu, Carl Richey, Tom Schnell, Frederic~D. McKenzie, and Jiang Li.
\newblock {EOG} artifact removal using a wavelet neural network.
\newblock \emph{Neurocomputing}, 97:\penalty0 374--389, November 2012.
\newblock \doi{10.1016/j.neucom.2012.04.016}.
\newblock URL \url{https://doi.org/10.1016/j.neucom.2012.04.016}.

\bibitem[Ghosh et~al.(2019)Ghosh, Sinha, and Biswas]{ghosh:2019:svm_autoencoder}
Rajdeep Ghosh, Nidul Sinha, and Saroj~Kumar Biswas.
\newblock Automated eye blink artefact removal from {EEG} using support vector machine and autoencoder.
\newblock \emph{{IET} Signal Processing}, 13\penalty0 (2):\penalty0 141--148, April 2019.
\newblock \doi{10.1049/iet-spr.2018.5111}.
\newblock URL \url{https://doi.org/10.1049/iet-spr.2018.5111}.

\bibitem[Yang et~al.(2018)Yang, Duan, Fan, Hu, and Wang]{yang:2018:dleegdenoise}
Banghua Yang, Kaiwen Duan, Chengcheng Fan, Chenxiao Hu, and Jinlong Wang.
\newblock Automatic ocular artifacts removal in {EEG} using deep learning.
\newblock \emph{Biomedical Signal Processing and Control}, 43:\penalty0 148--158, May 2018.
\newblock \doi{10.1016/j.bspc.2018.02.021}.
\newblock URL \url{https://doi.org/10.1016/j.bspc.2018.02.021}.

\bibitem[Sawangjai et~al.(2022)Sawangjai, Trakulruangroj, Boonnag, Piriyajitakonkij, Tripathy, Sudhawiyangkul, and Wilaiprasitporn]{swaangjai:2022:eeganet}
Phattarapong Sawangjai, Manatsanan Trakulruangroj, Chiraphat Boonnag, Maytus Piriyajitakonkij, Rajesh~Kumar Tripathy, Thapanun Sudhawiyangkul, and Theerawit Wilaiprasitporn.
\newblock {EEGANet}: Removal of ocular artifacts from the {EEG} signal using generative adversarial networks.
\newblock \emph{{IEEE} Journal of Biomedical and Health Informatics}, 26\penalty0 (10):\penalty0 4913--4924, October 2022.
\newblock \doi{10.1109/jbhi.2021.3131104}.
\newblock URL \url{https://doi.org/10.1109/jbhi.2021.3131104}.

\bibitem[Brophy et~al.(2022)Brophy, Redmond, Fleury, Vos, Boylan, and Ward]{brophy:2022:gandenoise}
Eoin Brophy, Peter Redmond, Andrew Fleury, Maarten~De Vos, Geraldine Boylan, and Tom{\'{a}}s Ward.
\newblock Denoising {EEG} signals for real-world {BCI} applications using {GANs}.
\newblock \emph{Frontiers in Neuroergonomics}, 2, January 2022.
\newblock \doi{10.3389/fnrgo.2021.805573}.
\newblock URL \url{https://doi.org/10.3389/fnrgo.2021.805573}.

\bibitem[Sun et~al.(2020)Sun, Su, Wu, and Wu]{sun:2020:dleegdenoise}
Weitong Sun, Yuping Su, Xia Wu, and Xiaojun Wu.
\newblock A novel end-to-end 1d-{ResCNN} model to remove artifact from {EEG} signals.
\newblock \emph{Neurocomputing}, 404:\penalty0 108--121, September 2020.
\newblock \doi{10.1016/j.neucom.2020.04.029}.
\newblock URL \url{https://doi.org/10.1016/j.neucom.2020.04.029}.

\bibitem[Zhang et~al.(2021{\natexlab{a}})Zhang, Zhao, Wei, Mantini, Li, and Liu]{zhang:2021:eegdenoisenet}
Haoming Zhang, Mingqi Zhao, Chen Wei, Dante Mantini, Zherui Li, and Quanying Liu.
\newblock {EEGdenoiseNet}: a benchmark dataset for deep learning solutions of {EEG} denoising.
\newblock \emph{Journal of Neural Engineering}, 18\penalty0 (5):\penalty0 056057, October 2021{\natexlab{a}}.
\newblock \doi{10.1088/1741-2552/ac2bf8}.
\newblock URL \url{https://doi.org/10.1088/1741-2552/ac2bf8}.

\bibitem[Yin et~al.(2022)Yin, Liu, Li, Qian, and Chen]{yin:2022:freqenhance_eegdenoise}
Jin Yin, Aiping Liu, Chang Li, Ruobing Qian, and Xun Chen.
\newblock Frequency information enhanced deep {EEG} denoising network for ocular artifact removal.
\newblock \emph{{IEEE} Sensors Journal}, pages 1--1, 2022.
\newblock \doi{10.1109/jsen.2022.3209805}.
\newblock URL \url{https://doi.org/10.1109/jsen.2022.3209805}.

\bibitem[Gao et~al.(2022)Gao, Chen, Tang, Ming, and Li]{Gao2022}
Tengfei Gao, Dan Chen, Yunbo Tang, Zhekai Ming, and Xiaoli Li.
\newblock {EEG} reconstruction with a dual-scale {CNN}-{LSTM} model for deep artifact removal.
\newblock \emph{{IEEE} Journal of Biomedical and Health Informatics}, pages 1--12, 2022.
\newblock \doi{10.1109/jbhi.2022.3227320}.
\newblock URL \url{https://doi.org/10.1109/jbhi.2022.3227320}.

\bibitem[Huang et~al.(2017)Huang, Liu, van~der Maaten, and Weinberger]{huang2017densely}
Gao Huang, Zhuang Liu, Laurens van~der Maaten, and Kilian~Q Weinberger.
\newblock Densely connected convolutional networks.
\newblock In \emph{Proceedings of the IEEE Conference on Computer Vision and Pattern Recognition}, 2017.

\bibitem[Klados and Bamidis(2016)]{klados:2016:eeg_eog_dataset}
Manousos~A. Klados and Panagiotis~D. Bamidis.
\newblock A semi-simulated {EEG}/{EOG} dataset for the comparison of {EOG} artifact rejection techniques.
\newblock \emph{Data in Brief}, 8:\penalty0 1004--1006, September 2016.
\newblock \doi{10.1016/j.dib.2016.06.032}.
\newblock URL \url{https://doi.org/10.1016/j.dib.2016.06.032}.

\bibitem[Mijovi{\'{c}} et~al.(2010)Mijovi{\'{c}}, Vos, Gligorijevi{\'{c}}, Taelman, and Huffel]{mijovic:2010:emd-ica}
Bogdan Mijovi{\'{c}}, M~De Vos, I~Gligorijevi{\'{c}}, J~Taelman, and S~Van Huffel.
\newblock Source separation from single-channel recordings by combining empirical-mode decomposition and independent component analysis.
\newblock \emph{{IEEE} Transactions on Biomedical Engineering}, 57\penalty0 (9):\penalty0 2188--2196, September 2010.
\newblock \doi{10.1109/tbme.2010.2051440}.
\newblock URL \url{https://doi.org/10.1109/tbme.2010.2051440}.

\bibitem[Zhang et~al.(2021{\natexlab{b}})Zhang, Wei, Zhao, Liu, and Wu]{zhang:2021:cnndenoise}
Haoming Zhang, Chen Wei, Mingqi Zhao, Quanying Liu, and Haiyan Wu.
\newblock A novel convolutional neural network model to remove muscle artifacts from {EEG}.
\newblock In \emph{{ICASSP} 2021 - 2021 {IEEE} International Conference on Acoustics, Speech and Signal Processing ({ICASSP})}. {IEEE}, June 2021{\natexlab{b}}.
\newblock \doi{10.1109/icassp39728.2021.9414228}.
\newblock URL \url{https://doi.org/10.1109/icassp39728.2021.9414228}.

\bibitem[Zhang et~al.(2022)Zhang, Yu, Rong, and Iwata]{zhang:2022:multimodule_denoise}
Zhen Zhang, Xiaoyan Yu, Xianwei Rong, and Makoto Iwata.
\newblock A novel multimodule neural network for {EEG} denoising.
\newblock \emph{{IEEE} Access}, 10:\penalty0 49528--49541, 2022.
\newblock \doi{10.1109/access.2022.3173261}.
\newblock URL \url{https://doi.org/10.1109/access.2022.3173261}.

\bibitem[Chuang et~al.(2022)Chuang, Chang, Huang, and Jung]{chuang:2022:icunet}
Chun-Hsiang Chuang, Kong-Yi Chang, Chih-Sheng Huang, and Tzyy-Ping Jung.
\newblock {IC}-u-net: A u-net-based denoising autoencoder using mixtures of independent components for automatic {EEG} artifact removal.
\newblock \emph{{NeuroImage}}, 263:\penalty0 119586, November 2022.
\newblock \doi{10.1016/j.neuroimage.2022.119586}.
\newblock URL \url{https://doi.org/10.1016/j.neuroimage.2022.119586}.

\bibitem[Wang et~al.(2020)Wang, Zou, Shen, and Ji]{wang:2020:non-local-unet}
Zhengyang Wang, Na~Zou, Dinggang Shen, and Shuiwang Ji.
\newblock Non-local u-nets for biomedical image segmentation.
\newblock In \emph{Proceedings of the AAAI Conference on Artificial Intelligence}, 2020.

\bibitem[Luo et~al.(2022)Luo, Zhang, Shu, Niu, and Zhou]{luo:2022:dlpr}
Wei Luo, Yi~Zhang, Xin Shu, Mengxuan Niu, and Renjie Zhou.
\newblock Learning end-to-end phase retrieval using only one interferogram with mixed-context network.
\newblock In Gabriel Popescu, YongKeun Park, and Yang Liu, editors, \emph{Quantitative Phase Imaging {VIII}}. {SPIE}, March 2022.
\newblock \doi{10.1117/12.2610502}.
\newblock URL \url{https://doi.org/10.1117/12.2610502}.

\bibitem[Ronneberger et~al.(2015)Ronneberger, Fischer, and Brox]{ronneberger:2015:unet}
Olaf Ronneberger, Philipp Fischer, and Thomas Brox.
\newblock U-net: Convolutional networks for biomedical image segmentation, 2015.

\bibitem[Yu and Koltun(2016)]{yu:2015:dilate}
Fisher Yu and Vladlen Koltun.
\newblock Multi-scale context aggregation by dilated convolutions.
\newblock In \emph{International Conference on Learning Representations}, 2016.

\bibitem[Tangermann et~al.(2012)Tangermann, M\"{u}ller, Aertsen, Birbaumer, Braun, Brunner, Leeb, Mehring, Miller, M\"{u}ller-Putz, Nolte, Pfurtscheller, Preissl, Schalk, Schl\"{o}gl, Vidaurre, Waldert, and Blankertz]{tangermann:2012:bci_competition}
Michael Tangermann, Klaus-Robert M\"{u}ller, Ad~Aertsen, Niels Birbaumer, Christoph Braun, Clemens Brunner, Robert Leeb, Carsten Mehring, Kai~J. Miller, Gernot~R. M\"{u}ller-Putz, Guido Nolte, Gert Pfurtscheller, Hubert Preissl, Gerwin Schalk, Alois Schl\"{o}gl, Carmen Vidaurre, Stephan Waldert, and Benjamin Blankertz.
\newblock Review of the bci competition iv.
\newblock \emph{Frontiers in Neuroscience}, 6, 2012.
\newblock ISSN 1662-4548.
\newblock \doi{10.3389/fnins.2012.00055}.
\newblock URL \url{http://dx.doi.org/10.3389/fnins.2012.00055}.

\bibitem[Chuang et~al.(2014)Chuang, Ko, Jung, and Lin]{chuang:2014:driving_task_dataset}
Chun-Hsiang Chuang, Li-Wei Ko, Tzyy-Ping Jung, and Chin-Teng Lin.
\newblock Kinesthesia in a sustained-attention driving task.
\newblock \emph{NeuroImage}, 91:\penalty0 187–202, May 2014.
\newblock ISSN 1053-8119.
\newblock \doi{10.1016/j.neuroimage.2014.01.015}.
\newblock URL \url{http://dx.doi.org/10.1016/j.neuroimage.2014.01.015}.

\bibitem[Ledig et~al.(2017)Ledig, Theis, Huszar, Caballero, Cunningham, Acosta, Aitken, Tejani, Totz, Wang, and Shi]{ledig:2017:srgan}
Christian Ledig, Lucas Theis, Ferenc Huszar, Jose Caballero, Andrew Cunningham, Alejandro Acosta, Andrew Aitken, Alykhan Tejani, Johannes Totz, Zehan Wang, and Wenzhe Shi.
\newblock Photo-realistic single image super-resolution using a generative adversarial network.
\newblock In \emph{2017 IEEE Conference on Computer Vision and Pattern Recognition (CVPR)}. IEEE, July 2017.
\newblock \doi{10.1109/cvpr.2017.19}.
\newblock URL \url{http://dx.doi.org/10.1109/CVPR.2017.19}.

\bibitem[Clark et~al.(1995)Clark, Biscay, Echeverría, and Virués]{clark:1995:eeg_multiresolution}
Ismael Clark, Rolando Biscay, Maribel Echeverría, and Trinidad Virués.
\newblock Multiresolution decomposition of non-stationary eeg signals: A preliminary study.
\newblock \emph{Computers in Biology and Medicine}, 25\penalty0 (4):\penalty0 373–382, July 1995.
\newblock ISSN 0010-4825.
\newblock \doi{10.1016/0010-4825(95)00014-u}.
\newblock URL \url{http://dx.doi.org/10.1016/0010-4825(95)00014-U}.

\bibitem[Vincent et~al.(2010)Vincent, Larochelle, Lajoie, Bengio, and Manzagol]{vincent:2010:ssae}
Pascal Vincent, Hugo Larochelle, Isabelle Lajoie, Yoshua Bengio, and Pierre-Antoine Manzagol.
\newblock Stacked denoising autoencoders: Learning useful representations in a deep network with a local denoising criterion.
\newblock \emph{J. Mach. Learn. Res.}, 11:\penalty0 3371–3408, dec 2010.
\newblock ISSN 1532-4435.

\bibitem[Margaux et~al.(2012)Margaux, Emmanuel, Sébastien, Olivier, and Jérémie]{margaux:2012:bci_ner}
Perrin Margaux, Maby Emmanuel, Daligault Sébastien, Bertrand Olivier, and Mattout Jérémie.
\newblock Objective and subjective evaluation of online error correction during p300-based spelling.
\newblock \emph{Advances in Human-Computer Interaction}, 2012:\penalty0 1–13, 2012.
\newblock ISSN 1687-5907.
\newblock \doi{10.1155/2012/578295}.
\newblock URL \url{http://dx.doi.org/10.1155/2012/578295}.

\bibitem[Jeong et~al.(2020)Jeong, Cho, Shim, Kwon, Lee, Lee, Lee, and Lee]{jeong:2020:multimodal_dataset}
Ji-Hoon Jeong, Jeong-Hyun Cho, Kyung-Hwan Shim, Byoung-Hee Kwon, Byeong-Hoo Lee, Do-Yeun Lee, Dae-Hyeok Lee, and Seong-Whan Lee.
\newblock Multimodal signal dataset for 11 intuitive movement tasks from single upper extremity during multiple recording sessions.
\newblock \emph{GigaScience}, 9\penalty0 (10), October 2020.
\newblock ISSN 2047-217X.
\newblock \doi{10.1093/gigascience/giaa098}.
\newblock URL \url{http://dx.doi.org/10.1093/gigascience/giaa098}.

\bibitem[Blankertz et~al.(2007)Blankertz, Dornhege, Krauledat, M\"{u}ller, and Curio]{blankertz:2007:bci_dataset}
Benjamin Blankertz, Guido Dornhege, Matthias Krauledat, Klaus-Robert M\"{u}ller, and Gabriel Curio.
\newblock The non-invasive berlin brain–computer interface: Fast acquisition of effective performance in untrained subjects.
\newblock \emph{NeuroImage}, 37\penalty0 (2):\penalty0 539–550, August 2007.
\newblock ISSN 1053-8119.
\newblock \doi{10.1016/j.neuroimage.2007.01.051}.
\newblock URL \url{http://dx.doi.org/10.1016/j.neuroimage.2007.01.051}.

\bibitem[Lee et~al.(2015)Lee, Xie, Gallagher, Zhang, and Tu]{lee:2015:deeply_supervised_nets}
Chen-Yu Lee, Saining Xie, Patrick Gallagher, Zhengyou Zhang, and Zhuowen Tu.
\newblock {Deeply-Supervised Nets}.
\newblock In Guy Lebanon and S.~V.~N. Vishwanathan, editors, \emph{Proceedings of the Eighteenth International Conference on Artificial Intelligence and Statistics}, volume~38 of \emph{Proceedings of Machine Learning Research}, pages 562--570, San Diego, California, USA, 09--12 May 2015. PMLR.
\newblock URL \url{https://proceedings.mlr.press/v38/lee15a.html}.

\bibitem[He et~al.(2015)He, Zhang, Ren, and Sun]{He2015}
Kaiming He, Xiangyu Zhang, Shaoqing Ren, and Jian Sun.
\newblock Delving deep into rectifiers: Surpassing human-level performance on {ImageNet} classification.
\newblock In \emph{2015 {IEEE} International Conference on Computer Vision ({ICCV})}. {IEEE}, December 2015.
\newblock \doi{10.1109/iccv.2015.123}.
\newblock URL \url{https://doi.org/10.1109/iccv.2015.123}.

\bibitem[Xu et~al.(2019)Xu, Zhang, and Xiao]{xu:2019:train_behaviour}
Zhi-Qin~John Xu, Yaoyu Zhang, and Yanyang Xiao.
\newblock Training behavior of deep neural network in frequency domain.
\newblock In \emph{Neural Information Processing}, pages 264--274. Springer International Publishing, 2019.
\newblock \doi{10.1007/978-3-030-36708-4_22}.
\newblock URL \url{https://doi.org/10.1007/978-3-030-36708-4_22}.

\bibitem[Elbert et~al.(1985)Elbert, Lutzenberger, Rockstroh, and Birbaumer]{elbert:1985:eog_denoise}
Thomas Elbert, Werner Lutzenberger, Brigitte Rockstroh, and Niels Birbaumer.
\newblock Removal of ocular artifacts from the {EEG} {\textemdash} a biophysical approach to the {EOG}.
\newblock \emph{Electroencephalography and Clinical Neurophysiology}, 60\penalty0 (5):\penalty0 455--463, May 1985.
\newblock \doi{10.1016/0013-4694(85)91020-x}.
\newblock URL \url{https://doi.org/10.1016/0013-4694(85)91020-x}.

\bibitem[Dora and Holcman(2022)]{dora:2022:adaptive_eegdenoise}
Matteo Dora and David Holcman.
\newblock Adaptive single-channel {EEG} artifact removal with applications to clinical monitoring.
\newblock \emph{{IEEE} Transactions on Neural Systems and Rehabilitation Engineering}, 30:\penalty0 286--295, 2022.
\newblock \doi{10.1109/tnsre.2022.3147072}.
\newblock URL \url{https://doi.org/10.1109/tnsre.2022.3147072}.

\bibitem[Barachant et~al.(2022)Barachant, Barthélemy, King, Gramfort, Chevallier, Rodrigues, Olivetti, Goncharenko, vom Berg, Reguig, Lebeurrier, Bjäreholt, Yamamoto, Clisson, and Corsi]{alexandre:2022:pyriemann}
Alexandre Barachant, Quentin Barthélemy, Jean-Rémi King, Alexandre Gramfort, Sylvain Chevallier, Pedro L.~C. Rodrigues, Emanuele Olivetti, Vladislav Goncharenko, Gabriel~Wagner vom Berg, Ghiles Reguig, Arthur Lebeurrier, Erik Bjäreholt, Maria~Sayu Yamamoto, Pierre Clisson, and Marie-Constance Corsi.
\newblock pyriemann/pyriemann: v0.3, July 2022.
\newblock URL \url{https://doi.org/10.5281/zenodo.7547583}.

\bibitem[W.Luo()]{xaikit}
W.Luo.
\newblock xai-kit.
\newblock URL \url{https://github.com/williamro/xai-kit}.

\bibitem[Jastrzebski et~al.(2018)Jastrzebski, Arpit, Ballas, Verma, Che, and Bengio]{jastrz:2018:residual_connection}
Stanisław Jastrzebski, Devansh Arpit, Nicolas Ballas, Vikas Verma, Tong Che, and Yoshua Bengio.
\newblock Residual connections encourage iterative inference.
\newblock In \emph{International Conference on Learning Representations}, 2018.
\newblock URL \url{https://openreview.net/forum?id=SJa9iHgAZ}.

\end{thebibliography}

\end{document}